%% file: TwistorLectures.tex
\documentclass[11pt, a4paper]{article}
\usepackage{jheparxiv}
\usepackage[utf8]{inputenc}
\usepackage{amsmath}
\usepackage{amsfonts}
\usepackage{amssymb}
\usepackage{latexsym}
\usepackage{mathrsfs}
\usepackage{graphicx}
\usepackage{color}
\usepackage{slashed}
\usepackage{twistor}
\usepackage[all]{xy}

\renewcommand{\d}{\mathrm{d}}

\title{Lectures on twistor theory}

\author{Tim Adamo}

\affiliation{Theoretical Physics Group, Blackett Laboratory \\
        Imperial College London, SW7 2AZ, United Kingdom}

\emailAdd{t.adamo@imperial.ac.uk}

\abstract{Broadly speaking, twistor theory is a framework for encoding physical information on space-time as geometric data on a complex projective space, known as a twistor space. The relationship between space-time and twistor space is non-local and has some surprising consequences, which we explore in these lectures. Starting with a review of the twistor correspondence for four-dimensional Minkowski space, we describe some of twistor theory's historic successes (e.g., describing free fields and integrable systems) as well as some of its historic shortcomings. We then discuss how in recent years many of these problems have been overcome, with a view to understanding how twistor theory is applied to the study of perturbative QFT today.

These lectures were given in 2017 at the XIII Modave Summer School in mathematical physics.}

\begin{document}
 
\maketitle

\setcounter{section}{-1}

\include{TLecture0}

\include{TLecture1}

\include{TLecture2}

\include{TLecture3}

\include{TLecture4}

\include{TLecture5}

\acknowledgments

I would like to thank the organizers of the XIII Modave Summer School in mathematical physics for inviting me to give these lectures and for putting together a very enjoyable school. These notes benefited greatly from the questions and comments of the participants at Modave; I would particularly like to thank Yannick Herfray, Tim de Jonckheere and Emanuel Malek in this regard. Many thanks to Alan Chodos for pointing out a typo in an earlier version. I am supported by an Imperial College Junior Research Fellowship.

\bibliography{TLectures}
\bibliographystyle{JHEP}
 
\end{document}

%% file: TLecture0.tex
\section{Introduction}

Twistor theory is a fascinating topic with a checkered past. It was first introduced fifty years ago by Penrose~\cite{Penrose:1967wn}, with the long-term ambition of developing a novel approach to quantum gravity. Despite many interesting initial advances, the subject stalled significantly by the late 1980s due to a variety technical and philosophical problems. For the following twenty years, twistor theory moved primarily into the realm of pure mathematics as a tool for the study of integrable systems and geometry. It was resurrected for physics in 2003 with Witten's observation~\cite{Witten:2003nn} (building on earlier work of Nair~\cite{Nair:1988bq}) that twistor theory can be combined with string perturbation theory to calculate the entire tree-level S-matrix of Yang-Mills theory in four space-time dimensions.

Today, twistor theory plays a prominent role in the study of interesting `non-standard' structures across a range of perturbative quantum field theories. Yet despite its wide applicability, twistor theory is not a subject that most graduate students in mathematical or theoretical physics are likely to encounter in their studies. The goal of these lectures is to provide graduate students (or more senior researchers who are encountering twistors for the first time) with an avenue into this vibrant and exciting arena of research.

As such, these lectures are not designed to be a painstaking exposition of the mathematical underpinnings of twistor theory. Nor are they meant to provide an introduction to the most cutting-edge aspects of research which make use of twistor methods. Rather, my hope is that after these lectures you will be able to look at any recent paper involving twistor theory (or some of its generalizations) and be able to understand the basics of what is happening. 

The intended audience are theoretical and mathematical physicists, rather than pure mathematicians. Thus, I have assumed a degree of familiarity with standard QFT notation and terminology, as well as a bit of general relativity. The final lecture assumes some exposure to the basics of string theory. Some background in mathematical subjects such as algebraic and differential geometry will make your life easier, but it is not essential: I have tried to provide basic (sometimes sketchy) explanations for all of the technical tools needed as they arise.

\medskip

In their original incarnation, these notes were delivered in five 1-hour lectures, but I expect that 90 minute lectures would be more suited to the presentation here. References throughout to the current research literature reflect my own interests and opinions, and are certainly incomplete. However, it would be useful to comment briefly on other pedagogical and reference treatments of twistor theory, since you will definitely want to refer to other sources if you are trying to learn the subject from scratch.

For my money, the best introductory textbook for twistor theory remains that of Huggett and Tod~\cite{Huggett:1985}; this book is well-written, covers all the basics, includes many exercises, and is remarkably compact. It would be my first recommendation to anyone who wants to learn enough twistor theory to get their hands dirty. 

The standard reference work in the subject is the two volume \textit{Spinors and space-time} by Penrose and Rindler~\cite{Penrose:1986,Penrose:1988}. This contains more-or-less everything that happened in twistor theory and related areas up to the late 1980s. The book \textit{Twistor Geometry and Field Theory} by Ward and Wells~\cite{Ward:1990} is also very useful, particularly for those approaching the subject from a mathematical background. Treatments more focused on the study of twistor theory and integrable systems are given by Mason and Woodhouse~\cite{Mason:1996rf} and Dunajski~\cite{Dunajski:2010zz}.

There have also been many review articles written about twistor theory over the years. One of the most cited is the \textit{Physics Reports} article by MacCallum and Penrose~\cite{Penrose:1972ia}; this serves as a useful introduction and includes many ideas that we will not have time to discuss in these lectures. The section on `The evaluation of scattering amplitudes' makes for particularly interesting reading in light of the modern development of the subject; you might understand why it took so long for twistor theory to make meaningful contact with the language of particle physics!

In the category of older review articles, the one by Woodhouse~\cite{Woodhouse:1985id} stands out as having aged particularly well. Its perspectives on many aspects of the subject are the ones used today, and much in these lectures makes use of Woodhouse's approach. More modern reviews, with a view towards applications in perturbative QFT can be found in~\cite{Jiang:2008xw,Adamo:2011pv,Adamo:2013cra}. The lecture course by Wolf~\cite{Wolf:2010av} provides an alternative exposition of many of the ideas presented in these lectures, as well as an introduction to the application of twistor theory to the study of scattering amplitudes in Yang-Mills theory. Finally, a recent historical overview of the subject was given by~\cite{Atiyah:2017erd}.

%% file: TLecture1.tex
\section{Spinor and Twistor Basics}\label{lect1}

We begin our study of twistor theory in the simple setting of flat, four-dimensional Minkowski space-time, $\M$, with signature $(+,-,-,-)$. Before jumping into twistor theory itself, it is important to set the stage using a few basic tools: complexification and spinor methods~\cite{Penrose:1986,Penrose:1988}. These will make our life substantially easier when talking about twistor theory, which is naturally defined for complexified space-time and phrased in terms of spinor variables. After this, we set out the basics of the twistor correspondence, focusing on the non-local relationship between twistor space and space-time.


\subsection{Complexified Minkowski space}

Let $\cM$ be a real, $d$-dimensional space-time equipped with a metric $\d s^2 = g_{ab}(x)\,\d x^{a}\,\d x^{b}$ in some coordinate system $x^a$. The \emph{complexification} of $(\cM, g_{ab})$ is defined by allowing the coordinates $x^{a}$ to take complex values while extending $g_{ab}(x)$ holomorphically~\cite{Penrose:1988}. Initially, each $x^{a}\in\R$ and the metric coefficients are real functions of the these real numbers; complexifying, we allow $x^{a}\in\C$ while the metric coefficients $g_{ab}(x)$ are now complex-valued, holomorphic functions of the $x^{a}$. (By `holomorphic,' we mean that there is no $\bar{x}^{a}$-dependence in the metric after complexification.) The resulting complexified space-time is denoted $\cM_{\C}$.

Let's focus on four-dimensional Minkowski space-time, $\M$. In Cartesian coordinates $x^{a}=(x^0,x^1,x^2,x^3)$, the metric is simply $\eta_{ab}=\mathrm{diag}(1,-1,-1,-1)$. Complexified Minkow-- ski space, $\M_{\C}$, is then just $\C^4$, equipped with the metric $\eta_{ab}$. The line element
\be\label{cMink1}
\d s^2=\eta_{ab}\,\d x^{a}\,\d x^{b}=(\d x^0)^2-(\d x^1)^2-(\d x^2)^2 - (\d x^3)^2\,,
\ee
looks the same as in real Minkowski space, with the exception that the coordinates are now allowed to take \emph{complex} values. 

Note that the `signature' of this complexified metric is no longer meaningful: real flat space of \emph{any} signature can be obtained by taking different real slices of the complexified space-time. The most obvious such real slice is that of real Minkowski space-time, $\M\subset\M_{\C}$. This corresponds to restricting the coordinates to take real values; in other words, just un-doing the process of complexification. However, by taking different real slices we can obtain $\R^4$ with \emph{Euclidean} signature $(+,+,+,+)$ or $\R^{2,2}$ with \emph{split} (or ultra-hyperbolic) signature $(+,+,-,-)$:
\begin{equation*}
 \mbox{Euclidean:}\:\: \R^4\subset\M_{\C}\,, \qquad x^{0}\in\R\,, \:\: x^{1},x^{2},x^{3}\in \im\R\,,
\end{equation*}
\begin{equation*}
 \mbox{Split:} \:\: \R^{2,2}\subset\M_{\C}\,, \qquad x^{0},x^{2},x^{3}\in\R\,, \:\: x^{1}\in\im\R\,.
\end{equation*}
In this sense, complexified Minkowski space is a sort of universal analytic continuation of all flat, real space-times.

Why do we care? Complexification means that we can study physics on $\M_{\C}$ (at least semi-classically), then recover results in the desired space-time signature by imposing appropriate reality conditions later. A calculation on $\M_{\C}$ will contain the corresponding calculations in any real space-time signature, provided we are careful about how we restrict to the real slice. This `moral' (i.e., `Complexify first, ask question later.') is a recurrent theme in twistor theory. Of course, at the end of the day we always want to wind up with real answers, so although later lectures often focus on calculations in the complexified setting and ignore the details of imposing reality conditions, we will spend some time in these early lectures emphasizing such details to make it clear how reality conditions are actually manifested.


\subsection{2-spinors in Minkowski space}

The spin group of complexified Minkowski space is SO$(4,\C)$, which is locally isomorphic to $\SL(2,\C)\times\SL(2,\C)$; in other words, the Lie algebra $\mathfrak{so}(4,\C)$ is isomorphic to $\mathfrak{sl}(2,\C)\times\mathfrak{sl}(2,\C)$.\footnote{This isomorphism is easy to see if you are familiar with the classification of semi-simple Lie algebras in terms of Dynkin diagrams.} A vector on $\M_{\C}$ lives in the $(\mathbf{\frac{1}{2}}, \mathbf{\frac{1}{2}})$ representation of $\SL(2,\C)\times\SL(2,\C)$, so any vector index can be represented by a pair of $\SL(2,\C)$ indices: one in the $(\mathbf{\frac{1}{2}},0)$ representation and the other in the $(0,\mathbf{\frac{1}{2}})$ representation.

The equivalence between a vector index on $\M_{\C}$ and two conjugate $\SL(2,\C)$ spinor indices is nothing to be afraid of: it is given by the familiar Pauli matrices, $\sigma_{a}$. Indeed, given a vector $v^{a}=(v^0,v^1,v^2,v^3)$, its representation in terms of $\SL(2,\C)$ Weyl spinors is given by:
\be\label{Pauli}
v^{\alpha\dot{\alpha}}:=\frac{\sigma_{a}^{\alpha\dot{\alpha}}}{\sqrt{2}}\,v^{a}=\frac{1}{\sqrt{2}}\left(\begin{array}{c c}
                                                                          v^{0}+v^{3} & v^{1}-\im v^{2} \\
                                                                          v^{1}+\im v^{2} & v^{0}-v^{3}
                                                                         \end{array}\right)\,.
\ee
The un-dotted spinor indices ($\alpha = 0,1$) live in the $(\mathbf{\frac{1}{2}},0)$ representation of $\SL(2,\C)\times\SL(2,\C)$, and will be referred to as \emph{negative chirality} spinor indices. The dotted spinor indices ($\dot{\alpha}=\dot{0},\dot{1}$) live in the $(0,\mathbf{\frac{1}{2}})$ representation and will be referred to as \emph{positive chirality} spinor indices. This rule (i.e., contracting with the Pauli matrices) allows us to replace any number of vector indices on $\M_{\C}$ with pairs of spinor indices. For instance, a rank-3 contravariant tensor $T^{abc}$ is translated into
\begin{equation*}
 T^{abc}\rightarrow T^{\alpha\dot{\alpha}\beta\dot{\beta}\gamma\dot{\gamma}}\,,
\end{equation*}
and so forth.

We can immediately observe one nice consequence of writing vectors in the 2-spinor formalism. Note that the norm of a vector $v^{a}$ with respect to the metric is encoded by the determinant of its spinor representation \eqref{Pauli}:
\be\label{Spinnorm}
\eta_{ab}\,v^{a}\,v^{b} = 2\,\det(v^{\alpha\dot{\alpha}})\,.
\ee
This means that $v^{a}$ is \emph{null} if and only if $\det(v^{\alpha\dot\alpha})$ vanishes. But $v^{\alpha\dot\alpha}$ is a $2\times2$ matrix, so its determinant vanishes if and only if its rank is less than two. Therefore, every (non-trivial) null vector in $\M_{\C}$ can be written as
\be\label{nullvect}
v^{\alpha\dot{\alpha}}_{\mathrm{null}}=a^{\alpha}\,\tilde{a}^{\dot{\alpha}}\,,
\ee
for some spinors $a^{\alpha},\tilde{a}^{\dot{\alpha}}$. The converse is also obviously true: any matrix of the form $a^{\alpha}\tilde{a}^{\dot{\alpha}}$ has vanishing determinant, and hence its corresponding vector is null. 

So the 2-spinor formalism provides an unconstrained way to represent null vectors in $\M_{\C}$: \emph{any} pair of Weyl spinors of opposite chirality define a null vector. This is certainly an improvement over the `standard' vectorial description, where one defines a null vector by specifying four (complex) numbers constrained by a quadratic equation.

\medskip

Of course, in order for it to be useful, we must be able to translate everything about the usual metric geometry of $\M_{\C}$ into the language of the 2-spinor formalism. In the standard language, we raise and lower indices using the metric tensor $\eta_{ab}$ or its inverse $\eta^{ab}$. The object we should used to raise and lower spinor indices are the natural $\SL(2,\C)$-invariant tensors, which are just the two-dimensional Levi-Civita symbols:
\be\label{Levi-Civita}
\epsilon_{\alpha\beta}=\left(\begin{array}{c c}
                              0 & 1 \\
                              -1 & 0
                             \end{array}\right) = \epsilon_{\dot{\alpha}\dot{\beta}}\,.
\ee
These objects are skew-symmetric ($\epsilon_{\alpha\beta}=-\epsilon_{\beta\alpha}$), and their inverses are defined by
\be\label{LC2}
\epsilon^{\alpha\beta}\,\epsilon_{\gamma\beta}=\delta^{\alpha}_{\beta}\,, \qquad \epsilon^{\alpha\beta}\,\epsilon_{\alpha\beta}=2\,,
\ee
and likewise for dotted indices.

Because they are skew-symmetric, it's important to fix a convention for how we raise and lower spinor indices and then stick to it -- otherwise, our calculations will be inconsistent due to sign errors. Our conventions will be `lower to the right, raise to the left':
\be\label{srlcon}
a_{\alpha}:=a^{\beta}\,\epsilon_{\beta\alpha}\,, \qquad b^{\alpha}:=\epsilon^{\alpha\beta}\,b_{\beta}\,,
\ee
with identical conventions for dotted (positive chirality) spinor indices. So given some vector $v^{\alpha\dot\alpha}$ (in spinor representation), this means that the dual covector is
\be\label{Pauli*}
v_{\alpha\dot{\alpha}}=v^{\beta\dot{\beta}}\,\epsilon_{\beta\alpha}\,\epsilon_{\dot{\beta}\dot{\alpha}}=\frac{1}{\sqrt{2}}\left(\begin{array}{c c}
                                                                                                               v^{0}-v^{3} & -(v^{1}+\im v^{2}) \\
                                                                                                               -v^{1}+\im v^{2} & v^{0}+v^{3}
                                                                                                              \end{array}\right)\,.
\ee
Sure enough, it is easy to see that $v^{\alpha\dot{\alpha}} v_{\alpha\dot{\alpha}}=2\det(v^{\alpha\dot{\alpha}})=\eta_{ab}v^{a}v^{b}$.  To summarize, in the 2-spinor formalism the line element for $\M_{\C}$ takes the form
\be\label{Minkmet}
\d s^{2}=\epsilon_{\alpha\beta}\,\epsilon_{\dot{\alpha}\dot{\beta}}\,\d x^{\alpha\dot{\alpha}}\,\d x^{\beta\dot\beta}\,,
\ee
where the coordinates $(x^0,x^1,x^2,x^3)$ take complex values and are encoded in the $2\times 2$ matrix $x^{\alpha\dot\alpha}$ according to \eqref{Pauli}.

\medskip

At this point, we will also introduce some notation which will make our lives easier as these lectures go along. Clearly, the Levi-Civita symbols define inner products on the spaces of negative and positive chirality spinors, respectively. We will denote these by:
\be\label{angsquare}
\la\kappa\,\omega\ra:=\kappa^{\alpha}\,\omega_{\alpha}=\kappa^{\alpha}\,\omega^{\beta}\,\epsilon_{\beta\alpha}\,, \qquad [\tilde{\kappa}\,\tilde{\omega}]:=\tilde{\kappa}^{\dot{\alpha}}\,\tilde{\omega}_{\dot{\alpha}}=\tilde{\kappa}^{\dot{\alpha}}\,\tilde{\omega}^{\dot{\beta}}\,\epsilon_{\dot{\beta}\dot{\alpha}}\,.
\ee
These are the natural $\SL(2,\C)$-invariant, skew-symmetric inner products on the 2-spinors of each chirality.

For example, consider any two null vectors $v^{a}_{\mathrm{null}}$ and $w^{a}_{\mathrm{null}}$ in $\M_{\C}$; as we noted above, these can be written as $v^{a}_{\mathrm{null}}\leftrightarrow\kappa^{\alpha}\tilde{\kappa}^{\dot{\alpha}}$ and $w^{a}_{\mathrm{null}}\leftrightarrow\omega^{\alpha}\tilde{\omega}^{\dot{\alpha}}$ for some spinors $\{\kappa^{\alpha},\tilde{\kappa}^{\dot{\alpha}},\omega^{\alpha},\tilde{\omega}^{\dot{\alpha}}\}$. The inner product of these two vectors is easily seen to be
\be\label{nvprod}
v_{\mathrm{null}}\cdot w_{\mathrm{null}} = \la\kappa\omega\ra\,[\tilde{\kappa}\tilde{\omega}]\,,
\ee
in terms of the inner products defined by \eqref{angsquare}.


\subsection{Real slices and spinor conjugations}

Having translated the metric geometry of $\M_{\C}$ into the language of 2-spinors, we now consider how real slices of various signature can be singled out at the level of the spinor formalism. This means finding reality conditions on the matrix
\be\label{coordmat}
x^{\alpha\dot\alpha}=\frac{1}{\sqrt{2}}\left(\begin{array}{c c}
                                              x^{0}+x^{3} & x^{1}-\im x^{2} \\
                                              x^{1}+\im x^{2} & x^{0}-x^{3}
                                             \end{array}\right)\,,
\ee
which are compatible with the desired signature. As we will see, each choice of reality condition induces a natural notion of `complex conjugation' on the spaces of spinors (c.f., \cite{Woodhouse:1985id}).

\subsubsection*{\textit{Lorentzian signature}}

Suppose we wish to single out the usual, Lorentzian real Minkowski space $\M$ inside of $\M_{\C}$. In terms of the usual coordinates $(x^0,x^1,x^2,x^3)$, we know that the appropriate reality condition is simply to force each of the $x^{a}$ to be real-valued. In terms of the matrix $x^{\alpha\dot{\alpha}}$, it is easy to see that this corresponds to requiring $x^{\alpha\dot{\alpha}}$ to be \emph{Hermitian}: $x^{\alpha\dot\alpha}=(x^{\alpha\dot{\alpha}})^{\dagger}$, where
\be\label{Lorreality}
(x^{\alpha\dot{\alpha}})^{\dagger}=\frac{1}{\sqrt{2}}\left(\begin{array}{c c}
                                              \bar{x}^{0}+\bar{x}^{3} & \bar{x}^{1}-\im \bar{x}^{2} \\
                                              \bar{x}^{1}+\im \bar{x}^{2} & \bar{x}^{0}-\bar{x}^{3}
                                             \end{array}\right)\,.
\ee
Since Hermitian conjugation includes the transpose operation (in addition to complex conjugation of the matrix entries), it is clear that positive and negative chirality spinor representations are exchanged when we compute $(x^{\alpha\dot{\alpha}})^\dagger$. Thus, the reality structure associated with the Lorentzian-real slice of $\M_{\C}$ is naturally associated with a complex conjugation on 2-spinors which exchanges dotted and un-dotted spinors.

In particular, given spinors with components $\kappa^{\alpha}=(a,b)$ and $\tilde{\omega}^{\dot{\alpha}}=(c,d)$, where $a,b,c,d\in\C$, the induced conjugation operation acts as:
\be\label{Lorconj}
\kappa^{\alpha}\mapsto \bar{\kappa}^{\dot{\alpha}}=(\bar{a},\bar{b})\,, \qquad \tilde{\omega}^{\dot\alpha}\mapsto\bar{\tilde{\omega}}^{\alpha}=(\bar{c},\bar{d})\,.
\ee
You can easily use this conjugation to show that any \emph{real} null vector in $\M$ can be written as $\kappa^{\alpha}\bar{\kappa}^{\dot\alpha}$ for some spinor $\kappa^{\alpha}$, and that this is compatible with the reality condition.

\subsubsection*{\textit{Euclidean signature}}

To fix the Euclidean real slice $\R^4$ inside $\M_{\C}$, define the following operation on $x^{\alpha\dot\alpha}$:
\be\label{Eucreal}
\hat{x}^{\alpha\dot\alpha}:=\frac{1}{\sqrt{2}}\left(\begin{array}{c c}
                                              \bar{x}^{0}-\bar{x}^{3} & -\bar{x}^{1}+\im \bar{x}^{2} \\
                                              -\bar{x}^{1}-\im \bar{x}^{2} & \bar{x}^{0}+\bar{x}^{3}
                                             \end{array}\right)\,.
\ee
Demanding that $x^{\alpha\dot\alpha}$ be preserved under this operation ($x^{\alpha\dot\alpha}=\hat{x}^{\alpha\dot\alpha}$) forces
\be\label{Eucreal2}
x^{\alpha\dot\alpha}|_{x=\hat{x}}=\frac{1}{\sqrt{2}}\left(\begin{array}{c c}
                                              x^{0}+\im y^{3} & \im y^{1}+ y^{2} \\
                                              \im y^{1}- y^{2} & x^{0}-\im y^{3}
                                              \end{array}\right)\,, \qquad x^{0},y^{1},y^{2},y^{3}\in\R\,.
\ee
It is easy to see that this is precisely the structure required to obtain the positive definite metric on $\R^4$: $x^2=2\det(x)=(x^0)^2+(y^1)^2+(y^2)^2+(y^3)^2$.

The `hat-operation' \eqref{Eucreal} induces a conjugation on 2-spinors which, unlike the Lor-- entzian conjugation, does not interchange spinor representations:
\be\label{Eucconj}
\kappa^{\alpha}\mapsto \hat{\kappa}^{\alpha}=(-\bar{b},\bar{a})\,, \qquad \tilde{\omega}^{\dot\alpha}\mapsto\tilde{\omega}^{\dot\alpha}=(-\bar{d},\bar{c})\,.
\ee
Note that this operation is qualitatively different from ordinary complex conjugation -- in particular, it does not square to the identity: $\hat{\hat{\kappa}}^{\alpha}=-\kappa^{\alpha}$. Indeed, we would need to apply the hat-conjugation \emph{four} times to get back to the spinor we started from. For this reason, the reality structure associated with Euclidean signature is often referred to as \emph{quaternionic}. 

One straightforward consequence of the quaternionic nature of the hat-conjugation acting on 2-spinors is that there is no non-trivial combination $\kappa^{\alpha}\tilde{\omega}^{\dot\alpha}$ which is preserved under the hat-operation. This is simply the statement that there are no real null vectors in Euclidean space!

\subsubsection*{\textit{Split signature}}

To fix the split signature real slice $\R^{2,2}$ inside $\M_{\C}$, we simply take the complex conjugate of $x^{\alpha\dot{\alpha}}$,
\be\label{Splreal}
\overline{x^{\alpha\dot\alpha}}=\frac{1}{\sqrt{2}}\left(\begin{array}{c c}
                                              \bar{x}^{0}+\bar{x}^{3} & \bar{x}^{1}+\im \bar{x}^{2} \\
                                              \bar{x}^{1}-\im \bar{x}^{2} & \bar{x}^{0}-\bar{x}^{3}
                                             \end{array}\right)\,,
\ee
and demand that $x^{\alpha\dot\alpha}=\overline{x^{\alpha\dot\alpha}}$. This forces
\be\label{Splreal2}
x^{\alpha\dot\alpha}|_{x=\bar{x}}=\frac{1}{\sqrt{2}}\left(\begin{array}{c c}
                                              x^{0}+x^{3} & x^{1}+y^{2} \\
                                              x^{1}- y^{2} & x^{0}-x^{3}
                                             \end{array}\right)\,, \qquad x^{0},x^{1},y^{2},x^{3}\in\R\,,
\ee
for which $x^2=2\det(x)=(x^0)^2+(y^2)^2-(x^1)^2-(x^3)^2$, as desired for split signature.

While the underlying conjugation on 2-spinors is ordinary complex conjugation, it does not interchange the spinor representations (since we simply took the complex conjugate of $x^{\alpha\dot\alpha}$ rather than the Hermitian conjugate). So in split signature the conjugation acts on spinors as:
\be\label{Splconj}
\kappa^{\alpha}\mapsto \overline{\kappa^{\alpha}}=(\bar{a},\bar{b})\,, \qquad \tilde{\omega}^{\dot\alpha}\mapsto\overline{\tilde{\omega}^{\dot\alpha}}=(\bar{c},\bar{d})\,.
\ee
Thus, 2-spinors on $\R^{2,2}$ are precisely those spinors whose components are \emph{real}-valued. In other words, the complexified spin group in split signature is simply $\SL(2,\R)\times\SL(2,\R)$. Any null vector on $\R^{2,2}$ can then be represented by $\kappa^{\alpha}\tilde{\kappa}^{\dot\alpha}$ for $\kappa^{\alpha}, \tilde{\kappa}^{\dot\alpha}\in\R^{2}$.


\subsection{Twistor space}

Having introduced the spinor formalism for complexified Minkowski space, we are now ready to define the twistor correspondence. Let $\CP^3$ be the 3-dimensional complex projective space: this is the space of all complex lines through the origin in $\C^{4}$. We can describe $\CP^3$ with homogeneous coordinates $Z^{A}=(Z^{1},Z^{2},Z^{3},Z^{4})$, which take values in the complex numbers, are never all vanishing, and are identified up to overall re-scalings:
\be\label{homogcoords}
(Z^{1},Z^{2},Z^{3},Z^{4})\neq(0,0,0,0)\,, \qquad r\,Z^{A}\sim Z^{A}\,, \quad \forall r\in\C^{*}\,,
\ee
where $\C^*$ is the set of all non-zero complex numbers. The invariance of the homogeneous coordinates under $\C^{*}$ rescalings (often called `projective' rescalings) means that the homogeneous coordinates only contain three (complex) degrees of freedom. In particular, we can chart $\CP^{3}$ by covering it with the coordinate patches $U_{i}=\{Z^{A}\in\C^{4}| Z^{i}\neq 0\}$; in $U_{i}$ there are manifestly three well-defined complex coordinates given by taking $(Z^{i})^{-1} Z^{A}$. For instance, on $U_{1}$ we have the coordinates $Z^{2}/Z^1$, $Z^3/Z^1$, and $Z^4/Z^1$. 

The \emph{twistor space} $\PT$ of complexified Minkowski space is defined to be an open subset of the complex projective space $\CP^3$. In the next lecture, we'll learn exactly \emph{which} open subset we should choose, but for now this is not important. On $\PT$ it is useful to divide the four homogeneous coordinates $Z^A$ into two Weyl spinors of opposite chirality:
\be\label{twistcoords}
Z^{A}=(\mu^{\dot\alpha},\,\lambda_{\alpha})\,,
\ee
where $\mu^{\dot\alpha}$ and $\lambda_{\alpha}$ carry the same weight with respect to projective rescalings. In other words, the division of the $Z^A$ into $\mu^{\dot\alpha}$ and $\lambda_{\alpha}$ is nothing but fancy notation at this point.

The non-trivial step is defining a relationship between $\PT$ and space-time. This relationship is non-local, and is often referred to as the \emph{twistor correspondence}. For complexified Minkowski space, the twistor correspondence is captured by an algebraic relation between the coordinates $Z^A$ on twistor space and the coordinates $x^{\alpha\dot\alpha}$ on $\M_{\C}$:
\be\label{increls}
\mu^{\dot\alpha}=x^{\alpha\dot\alpha}\,\lambda_{\alpha}\,.
\ee
These equations are known as the \emph{incidence relations} -- they are the root of everything that is interesting about twistor theory.

In more formal treatments of twistor theory, this relationship is often presented in terms of a \emph{double fibration} of the projective spinor bundle over $\M_{\C}$ and $\PT$:
\begin{equation*}
\xymatrix{
 & \PS \ar[ld]_{\pi_{2}} \ar[rd]^{\pi_{1}} & \\
 \PT & & \M_{\C}} 
\end{equation*}
where $\PS$ has coordinates $(x^{\alpha\dot\alpha},\lambda_{\beta})$, with $\lambda_{\beta}\sim r\lambda_{\beta}$ for all non-zero complex numbers $r$. This means that on $\PS$, the spinor $\lambda_\beta$ acts as a homogeneous coordinate on the one-dimensional complex projective space $\CP^1$, which is just the Riemann sphere. So $\PS\cong\M_{\C}\times\CP^1$, and the map $\pi_{1}:\PS\rightarrow \M_{\C}$ is simply the projection $(x^{\alpha\dot\alpha},\lambda_\beta)\mapsto x^{\alpha\dot\alpha}$ while $\pi_{2}:\PS\rightarrow\PT$ imposes the incidence relations, $(x^{\alpha\dot\alpha},\lambda_\beta)\mapsto (x^{\beta\dot\alpha}\lambda_{\beta}, \lambda_{\alpha})$.

For our purposes, it suffices to think about twistor space purely in terms of the incidence relations \eqref{increls}, though. What do these relations actually tell us? First of all, suppose that we fix a point $x\in\M_{\C}$; what does this correspond to in twistor space? From \eqref{increls}, $x^{\alpha\dot\alpha}$ are coefficients in a linear equation relating $\mu^{\dot\alpha}$ and $\lambda_{\alpha}$. Suppose that we forgot about the projective scale of the coordinates on twistor space for a moment; then $Z^{A}=(\mu^{\dot\alpha}, \lambda_{\alpha})$ are just coordinates on $\C^4$ and the incidence relations $\mu^{\dot\alpha}=x^{\alpha\dot\alpha}\lambda_\alpha$ define a complex plane $\C^2\subset\C^4$. Putting the projective scale back into the game, we find that the incidence relations (for fixed $x^{\alpha\dot\alpha}$) define a $\CP^1\subset\PT$. Since the equation is linear and holomorphic (i.e., there are no complex conjugations appearing anywhere), it seems that a point in $\M_{\C}$ corresponds to a linearly and holomorphically embedded Riemann sphere in twistor space.

We can be even more precise about this: \emph{any} holomorphic linear embedding of a Riemann sphere into $\CP^3$ (or an open subset thereof) can always be put into the form of the incidence relations for fixed $x^{\alpha\dot\alpha}$. If $\sigma_{\mathrm{a}}=(\sigma_0, \sigma_1)$ are homogeneous coordinates on $\CP^1$, then such a map is given by
\be\label{holomap1}
\mu^{\dot\alpha}=b^{\dot{\alpha}\mathrm{a}}\,\sigma_{\mathrm{a}}\,, \qquad \lambda_{\alpha}=c^{\mathrm{a}}_{\alpha}\,\sigma_{\mathrm{a}}\,,
\ee
where the 8 complex parameters $(b^{\dot{\alpha}\mathrm{a}},c^{\mathrm{a}}_{\alpha})$ define the map. Of course, this is over-counting: we haven't taken into account the automorphism group of the Riemann sphere or the projective rescalings of the homogeneous coordinates of the $\CP^3$ target space. This is 4 complex degrees of freedom (3 from the automorphisms of $\CP^1$, which are the M\"obius transformations, and 1 for the $\C^*$ projective rescalings), which can be used to fix $c^{\mathrm{a}}_{\alpha}=\delta^{\mathrm{a}}_{\alpha}$. After fixing this redundancy in \eqref{holomap1}, the map looks like
\be\label{holomap2}
 \mu^{\dot\alpha}=b^{\dot{\alpha}\mathrm{a}}\,\sigma_{\mathrm{a}}\,, \qquad \lambda_{\alpha}=\delta^{\mathrm{a}}_{\alpha}\,\sigma_{\mathrm{a}}\,,
\ee
which is precisely the incidence relations \eqref{increls} with $x^{\alpha\dot\alpha}$ identified with $b^{\dot{\alpha}\mathrm{a}}$.

The upshot of this is that a point in Minkowski space corresponds to a holomorphically, linearly embedded Riemann sphere in twistor space. For a point $x\in\M_{\C}$, we denote the corresponding Riemann sphere in twistor space by $X\cong\CP^1\subset\PT$. We will often refer to these Riemann spheres as `lines' (e.g., `The line $X$ associated to $x\in\M_{\C}$.'), since they are linearly embedded and defined holomorphically. This is our first taste of the non-locality of the relationship between $\PT$ and $\M_{\C}$: a point in space-time is described by an extended object in twistor space!

\medskip

What about the other way around? That is, what does a point in twistor space correspond to in space-time? To answer this question, it is illuminating to describe a point $Z\in\PT$ as the intersection of two lines (that is, holomorphic, linearly embedded Riemann spheres), say $X$ and $Y$. By the incidence relations, this means that
\be\label{intlines}
X\cap Y=\{Z\in\PT\} \, \Rightarrow \quad \mu^{\dot{\alpha}}=x^{\alpha\dot\alpha}\,\lambda_{\alpha}\,\:\:\:\mathrm{and}\:\:\:\mu^{\dot{\alpha}}=y^{\alpha\dot\alpha}\,\lambda_{\alpha}\,,
\ee
for two points $x,y\in\M_{\C}$. Subtracting one incidence relation from the other, we discover that
\be\label{intlines2}
(x-y)^{\alpha\dot{\alpha}}\,\lambda_{\alpha}=0\,.
\ee
In this equation, contraction on the undotted spinor index is accomplished through the anti-symmetric $\epsilon_{\alpha\beta}$; since this is a 2-dimensional object, the only way that \eqref{intlines2} can hold (without $(x-y)^{\alpha\dot{\alpha}}$ being zero) is if $(x-y)^{\alpha\dot\alpha}\propto\lambda^{\alpha}$.

Therefore, the lines $X,Y$ in twistor space intersect in a point $Z$ if and only if their difference obeys
\be\label{intlines3}
(x-y)^{\alpha\dot\alpha}=\lambda^{\alpha}\,\tilde{\lambda}^{\dot\alpha}\,,
\ee
for some $\tilde{\lambda}^{\dot\alpha}$. But this means that $x,y\in\M_{\C}$ are null separated! So we discover that lines in twistor space intersect if and only if their corresponding points in $\M_{\C}$ are null separated. The point $Z\in\PT$ is described in $\M_{\C}$ by varying over the choice of the spinor $\tilde{\lambda}^{\dot\alpha}$ in \eqref{intlines3}. The result is a 2-plane (because there are two degrees of freedom in $\tilde{\lambda}^{\dot\alpha}$) which is totally null: every tangent vector to the plane is of the form $\lambda^{\alpha}\tilde{\lambda}^{\dot\alpha}$, where $\lambda^{\alpha}$ is fixed by the undotted components of $Z\in\PT$. These planes are referred to as $\alpha$-\emph{planes}.

\begin{figure}[t]
\centering
\includegraphics[width=80mm]{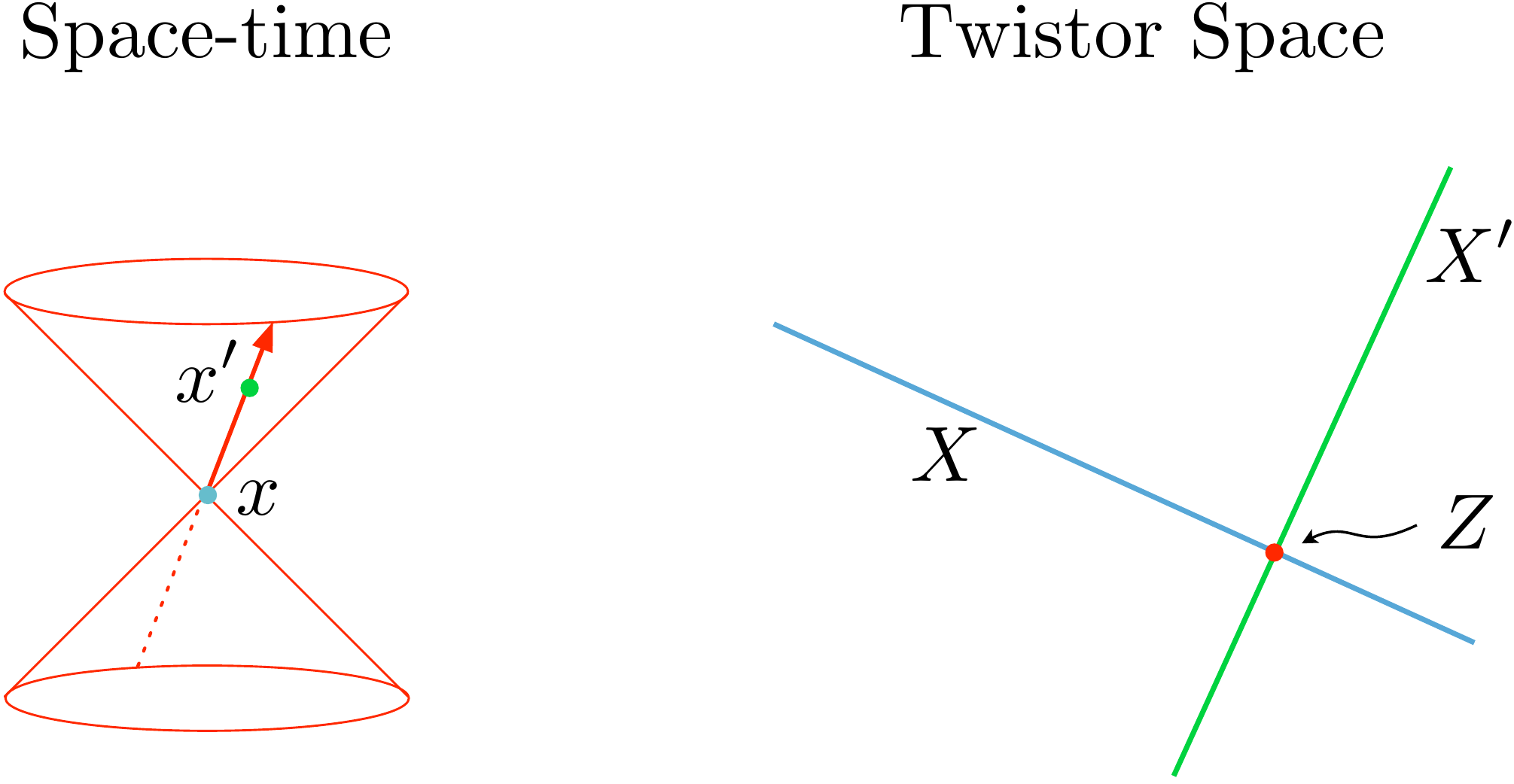}
\caption{\emph{The geometry of the twistor correspondence.}}
\label{tcorr}
\end{figure}

So the non-locality of the twistor correspondence is manifest in both directions: a point in twistor space corresponds to an $\alpha$-plane in $\M_{\C}$, while a point in $\M_{\C}$ corresponds to a linearly embedded Riemann sphere in twistor space; see Figure~\ref{tcorr}. Furthermore, the correspondence captures the conformal structure of (complexified) space-time, since points lying on the light cone of $x\in\M_{\C}$ are uniquely identified in twistor space by the lines which intersect $X\subset\PT$. The correspondence is also stated in purely holomorphic terms on twistor space, which brings us to a moral of twistor theory: holomorphic structures on twistor space encode conformal structures on space-time.

\medskip

\subsubsection*{Exercise: \textit{Points in $\M_\C$ as bi-twistors}}

We've learned that a point in space-time is represented in twistor space by a linearly embedded Riemann sphere, or line, $X$. Just like a line in three real dimensions is specified by any two points which lie on that line, so a holomorphic line in 3 complex dimensions is uniquely specified by any two points which lie on that line. Let $Z_{1},Z_{2}$ be two points in $\PT$ which lie on the line $X$. This means we can represent the line by taking the skew product of these two points, $Z_{1}\wedge Z_{2}$.

Using the incidence relations, show that the resulting `bi-twistor' $X^{AB}=Z_{1}^{[A}Z_{2}^{B]}$ takes the form:
\be\label{bitwistor}
X^{AB}=\la\lambda_{1}\,\lambda_{2}\ra\,  \left(\begin{array}{c c}
                                                                            \frac{1}{2}\epsilon^{\dot{\alpha}\dot{\beta}} x^{2} & x^{\dot\alpha}_{\beta} \\
                                                                            -x^{\dot\beta}_{\alpha} & \epsilon_{\alpha\beta}
                                                                            \end{array}\right)\,.
\ee
In particular, the skew bi-twistor encodes precisely the information of the space-time point $x^{\alpha\dot\alpha}$ up to a scale set by $\la\lambda_{1}\,\lambda_{2}\ra$.

%% file: TLecture2.tex
\section{Twistor Geometry}

We have seen that twistor space is related non-locally to complexified Minkowski space: points in space-time correspond to holomorphic, linearly embedded Riemann spheres (or `lines') in twistor space. The conformal structure of space-time is encoded by the holomorphic structure of these lines in twistor space: lines intersect if and only if the corresponding space-time points are null separated. In this lecture, we explore further how structures on space-time (in particular, reality structures and conformal structures) are translated into geometric structures on twistor space.


\subsection{Reality structures}

In the previous lecture, we discussed how the various real signature slices of $\M_{\C}$ can be recovered by imposing reality conditions. In the 2-spinor formalism, these reality conditions induced notions of complex conjugation on the spaces of spinors. Our goal is now to understand how these reality conditions are translated into twistor space. In other words, what conditions do we need to impose on $\PT$ (an open subset of $\CP^3$) so that it is related to a particular real slice of $\M_\C$ under the twistor correspondence?

\subsubsection*{\textit{Lorentzian signature}}

For real Minkowski space $\M$, recall that the natural conjugation on 2-spinors is the ordinary complex conjugation with the proviso that the positive (dotted) and negative (un-dotted) chirality spinor representations are exchanged under the conjugation. So given a twistor $Z^{A}=(\mu^{\dot\alpha},\lambda_{\alpha})$, the complex conjugation acts on the components as
\be\label{LorPT1}
(\mu^{\dot\alpha},\,\lambda_{\alpha})\mapsto (\bar{\lambda}_{\dot\alpha},\,\bar{\mu}^{\alpha})\,.
\ee
Thus, the complex conjugation naturally sends a twistor to something with its component indices in complimentary representations. There is a natural way to interpret this in terms of a `duality' on twistor space (this is actually an example of something known as projective duality).

To make our lives easier, in Lorentzian signature we modify the incidence relations \eqref{increls} by including a factor of `$\im$':
\be\label{lorincrels}
\mu^{\dot\alpha}=\im\,x^{\alpha\dot\alpha}\,\lambda_{\alpha}\,.
\ee
The geometry of the basic twistor correspondence is completely unchanged by this modification, and we only work with \eqref{lorincrels} in the specific context of Lorentzian reality conditions. Let $\PT^{\vee}$ be the same open subset of $\CP^3$ as $\PT$, but now with homogeneous coordinates $W_{A}=(\tilde{\lambda}_{\dot\alpha},\tilde{\mu}^{\alpha})$. Points in this \emph{dual twistor space} are related to points in $\M_{\C}$ by incidence relations:
\be\label{dualincrels}
\tilde{\mu}^{\alpha}=-\im\,x^{\alpha\dot\alpha}\,\tilde{\lambda}_{\dot\alpha}\,.
\ee
There is a natural inner product between $\PT$ and $\PT^{\vee}$ given by contracting a twistor index against a dual twistor index
\be\label{dualip}
Z\cdot W:=Z^{A}\,W_{A}=[\mu\,\tilde{\lambda}] + \la \tilde{\mu}\,\lambda\ra\,,
\ee
in terms of the $\SL(2,\C)$-invariant inner products on dotted and undotted spinors.

Coming back to the Lorentzian reality structure, we can now say that the complex conjugation maps a twistor $Z^{A}$ to a point in dual twistor space, $\bar{Z}_{A}$, whose components are the complex conjugates of the original twistor. Thus, complex conjugation induces an inner product on twistor space of the form
\be\label{tip}
Z\cdot\bar{Z}=[\mu\,\bar{\lambda}] + \la\bar{\mu}\,\lambda\ra\,.
\ee
Using the anti-symmetry of the spinor inner products, we see that \eqref{tip} has signature $(+2,-2)$ when viewed non-projectively (that is, as an inner product on $\C^4$). Since Lorent-- zian-real Weyl spinors are valued in $\SU(2)$, this means that the inner product is an $\SU(2,2)$-invariant. So the spinor conjugation appropriate to Lorentzian Minkowski space $\M$ induces a degenerate, $\SU(2,2)$-invariant inner product on twistor space.

Given a line $X\cong\CP^1$ in $\PT$, how do we know that the corresponding space-time point $x^{\alpha\dot\alpha}$ is valued in the real Minkowski space? Let $Z\in X$ be any point lying on the line in twistor space. Using the incidence relations, it follows that
\be\label{LorPT2}
Z\cdot\bar{Z}=\im\,x^{\alpha\dot\alpha}\,\lambda_{\alpha}\,\bar{\lambda}_{\dot\alpha}-\im(x^{\alpha\dot\alpha})^{\dagger}\,\bar{\lambda}_{\dot\alpha}\,\lambda_{\alpha}=\im\,(x-x^{\dagger})^{\alpha\dot\alpha}\,\lambda_{\alpha}\,\bar{\lambda}_{\dot\alpha}\,.
\ee
But we know that $x\in\M$ if and only if $x^{\alpha\dot\alpha}=(x^{\alpha\dot\alpha})^{\dagger}$. Therefore, any line $X$ which corresponds to a point in real Minkowski space-time must be contained in
\be\label{LorPT3}
\PN=\left\{Z\in\PT | Z\cdot\bar{Z}=0\right\}\,.
\ee
In other words, $\PN\subset\PT$ is the twistor space associated with $\M$; in the twistor theory literature $\PN$ is often referred to as the `space of null twistors.'

\begin{figure}[t]
\centering
\includegraphics[width=60mm]{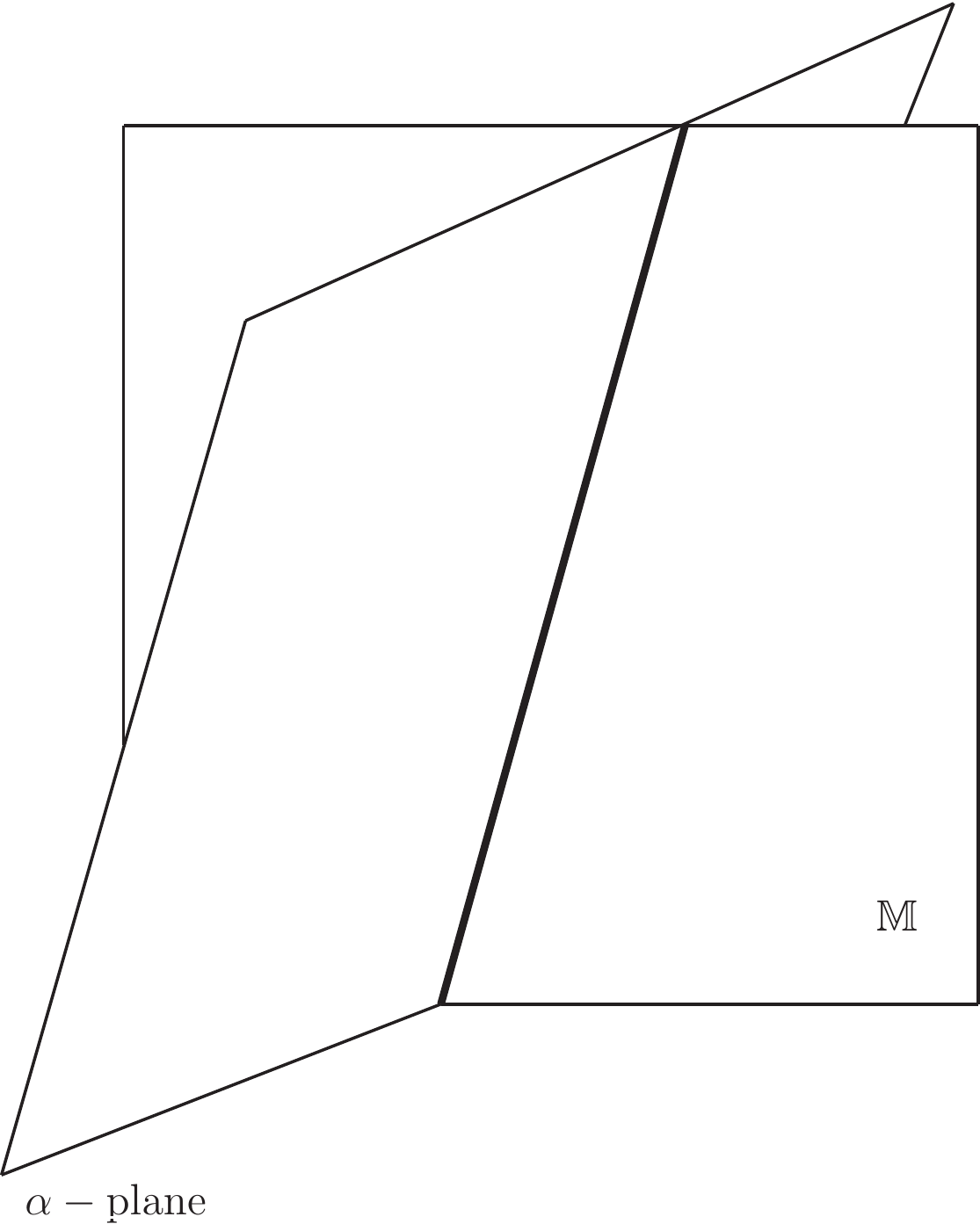}
\caption{\emph{The intersection between the Lorentzian real slice $\M\subset\M_{\C}$ and the $\alpha$-plane of a twistor $Z\in\PN$ is a real null geodesic.}}
\label{Reality1}
\end{figure}

Recall that a point in $\PT$ corresponds to an $\alpha$-plane -- a totally null complex 2-plane whose tangent vectors are all proportional to $\lambda_{\alpha}$, the un-dotted spinor components of $Z^A$ -- in complexified Minkowski space. What does a point in $\PN$ correspond to in $\M_{\C}$? You can show that the condition $Z\cdot\bar{Z}=0$ singles out a single tangent vector to $\alpha$-plane, namely: $\lambda^{\alpha}\bar{\lambda}^{\dot\alpha}$. Thus, a point $Z\in\PN$ corresponds to a unique real null geodesic, $\lambda^{\alpha}\bar{\lambda}^{\dot\alpha}$, in $\M$. The picture is that this real null geodesic is where the complex $\alpha$-plane intersects the real slice $\M$ of $\M_{\C}$; see Figure~\ref{Reality1}. Lines in $\PN$ intersect if and only if their corresponding points in $\M$ are separated by such a real null geodesic.

\subsubsection*{\textit{Euclidean signature}}

The reality structure associated with Euclidean $\R^4$ inside of $\M_{\C}$ induced a \emph{quaternionic} conjugation on spinors, which acts as
\be\label{EuPT1}
\kappa^{\alpha}=(a,b)\mapsto \hat{\kappa}^{\alpha}=(-\bar{b},\bar{a})\,, \qquad \tilde{\omega}^{\dot\alpha}=(c,d)\mapsto\tilde{\omega}^{\dot\alpha}=(-\bar{d},\bar{c})\,.
\ee
Acting on twistor space, this conjugation induces an involution $\sigma:\PT\rightarrow\PT$ sending
\be\label{EuPT2}
Z^{A}=(\mu^{\dot\alpha},\,\lambda_{\alpha})\mapsto \hat{Z}^{A}=(\hat{\mu}^{\dot\alpha},\,\hat{\lambda}_{\alpha})\,.
\ee
Since $\sigma^2=-\mathrm{id}$, it is clear that there are no points in twistor space which are preserved with respect to this conjugation. This makes sense: a point in $\PT$ is a totally null $\alpha$-plane in $\M_\C$, and the statement that there are no real (with respect to $\sigma$) points in $\PT$ is the statement that this $\alpha$-plane does not intersect the real slice $\R^4$, or that there are no real null geodesics in positive-definite signature.

Even if there are no real points in $\PT$, we can still ask if there are \emph{lines} which are preserved by $\sigma$. You (hopefully) showed that any line $X$ in twistor space can be represented by a skew bi-twistor $X^{AB}=Z_{1}^{[A}Z_{2}^{B]}$, where $Z_1, Z_2$ are any two distinct points lying on $X$. Clearly, any line of the form $X^{AB}=Z^{[A}\hat{Z}^{B]}$ will be preserved, since $\hat{X}^{AB}=X^{AB}$. This means that with Euclidean reality conditions, every point $Z\in\PT$ is uniquely associated with a point $x\in\R^4$ by taking the line passing through $Z$ and its conjugate $\hat{Z}$: $X^{AB}=Z^{[A}\hat{Z}^{B]}$.  

The fancy way of saying this is that Euclidean reality conditions induce a $\CP^1$ fibration $\PT\rightarrow\R^4$: every point of twistor space gets mapped to a point of $\R^4$ using the reality conditions, while every point of $\R^4$ corresponds to a $\CP^1$ worth of points (the twistor line $X$) in twistor space. At the level of spinor variables, this fibration is given explicitly by
\be\label{EuPT3}
x^{\alpha\dot\alpha}=\frac{\hat{\mu}^{\dot\alpha}\lambda^{\alpha}-\mu^{\dot\alpha}\hat{\lambda}^{\alpha}}{\la\lambda\,\hat{\lambda}\ra}\,.
\ee
It is easy to see that this is real with respect to the quaternionic conjugation and is compatible with the incidence relations in the sense that $x^{\alpha\dot\alpha}\lambda_{\alpha}=\mu^{\dot\alpha}$. 

So in Euclidean signature, a point in twistor space can be specified by fixing a point in $\R^4$ (i.e., a line which is preserved by $\sigma$) and then a point on the corresponding Riemann sphere. In other words the Euclidean twistor space is isomorphic to $\R^4\times\CP^1$ with coordinates $(x^{\alpha\dot\alpha},\lambda_{\alpha})$. This means that Euclidean reality conditions identify the twistor space of $\R^4$ with the projective spinor bundle $\PS\cong\R^4\times\CP^1$. Although points of twistor space are mapped to points of $\R^4$, the twistor correspondence remains non-local since a full Riemann sphere in twistor space corresponds to the same point on $\R^4$.

\subsubsection*{\textit{Split signature}}

For the real slice $\R^{2,2}$, we saw that the appropriate conjugation on 2-spinors was ordinary complex conjugation which does \emph{not} exchange spinor representations. In other words, 2-spinors of $\R^{2,2}$ are manifestly real $\SL(2,\R)$ spinors. This complex conjugation acts as an involution on twistor space,
\be\label{SpPT1}
Z^{A}=(\mu^{\dot\alpha},\,\lambda_{\alpha})\mapsto \overline{Z^A}=(\overline{\mu^{\dot\alpha}},\,\overline{\lambda_\alpha})\,.
\ee
So the natural portion of twistor space which is preserved by this complex conjugation is formed by the points of $\PT$ which are (literally) real-valued: $\PT_{\R}\subset\RP^3$. 

It is easy to see that $\PT_\R$ is the correct twistor space for $\R^{2,2}$. Take a line $X\subset \PT_\R$; then for any point $Z\in X$ it follows that $Z=\bar{Z}$ and thus the incidence relations imply that
\be\label{SpPT2}
(x-\overline{x})^{\alpha\dot\alpha}\,\lambda_{\alpha}=0\,.
\ee
But $x^{\alpha\dot\alpha}=\overline{x^{\alpha\dot\alpha}}$ for points in $\R^{2,2}$, so the equation is trivially satisfied. Hence, the twistor theory of split signature Minkowski space is a theory of real variables.

\medskip

In general, the idea in twistor theory is to work in the complexified setting, imposing reality conditions only at the end of a calculation. In the old days of the subject, these reality conditions were usually the Lorentzian ones, while early in the `twistor renaissance' of 2004 the split signature reality conditions were preferred. Nowadays, Euclidean reality conditions seem to be the most useful when performing explicit calculations. So depending on what era of the literature you read, you can find any one of the three reality conditions given preference for a combination of physical and technical reasons. In these lectures, we will focus mainly on Euclidean signature, for the following reasons: it maintains the complex-projective features of the general complexified signature twistor correspondence (unlike split signature); it has the nice feature that twistor space is a $\CP^1$-bundle over space-time in Euclidean signature; and many of the recent applications of twistor theory to the study of perturbative QFT are most cleanly phrased in these reality conditions.

\subsection{Complex structures}

Recall that one of the `morals' of twistor theory is that a complex structure on $\PT$ determines a conformal structure on space-time and \emph{vice versa}. This is manifest already in the basic geometry of the twistor correspondence: the conformal structure of $\M_\C$ is determined by the intersections of holomorphic lines in twistor space. What exactly is a complex structure on twistor space? Intuitively, we have described it as a way of knowing when things (e.g., functions, vectors, etc.) are holomorphic. 

If you've had a course on complex geometry, you will have heard that an almost complex structure on a manifold $M$ is a linear map $J:TM\rightarrow TM$ on the tangent bundle $TM$ of the manifold which obeys $J^2=-\mathrm{id}$. In component notation, if $i,j,\ldots$ are vector/covector indices on $M$, then the almost complex structure is a rank-two tensor $J^{i}_{j}$ which maps a vector $V^{i}$ to $J^{i}_{j}V^{j}$ and has the property $J^{i}_{j} J^{j}_{k}=-\delta^{i}_{k}$. To each $J$, we can associate an object called the \emph{Nijenhuis tensor}, $N_{J}$, which you should think of as a sort of curvature associated with the almost complex structure. In local coordinates, it is given by
\be\label{Nijenhuis}
(N_{J})^{k}_{ij}=J^{l}_{j}\,\partial_{l}J^{k}_{i}-J^{l}_{i}\,\partial_{l}J^{k}_{j} + J^{k}_{l}\left(\partial_{i}J^{l}_{j}-\partial_{j}J^{l}_{i}\right)\,.
\ee
An almost complex structure is said to be \emph{integrable} (i.e., a complex structure) if $N_J=0$.

We will adopt a slightly different, but equivalent, perspective on almost complex structures. Since $J$ is an isomorphism of $TM$ which squares to $J^2=-\mathrm{id}$, the complexified tangent bundle $TM_{\C}$ can be decomposed into eigenspaces of $J$ with eigenvalues $+\im$ and $-\im$. Vector fields with eigenvalue $+\im$ under $J$ are referred to as `holomorphic vector fields', or $(1,0)$-vectors, and those with eigenvalue $-\im$ are referred to as `anti-holomorphic vector fields' or $(0,1)$-vectors:
\be\label{tbdecomp}
TM_{\C}=T^{(1,0)}_{M}\oplus T^{(0,1)}_{M}\,.
\ee
If $(z^{a},\bar{z}^{\bar{a}})$ are local complex coordinates on $M$, then this decomposition is simply
\begin{equation*}
 V^{i}\frac{\partial}{\partial x^i} = V^{a}\frac{\partial}{\partial z^{a}} \oplus V^{\bar{a}}\frac{\partial}{\partial\bar{z}^{\bar{a}}}\,,
\end{equation*}
in terms of the local coordinates.

This decomposition naturally extends to differential forms on $M$: the decomposition for 1-forms, or covectors, is induced from \eqref{tbdecomp} by the natural pairing between vectors and 1-forms, and this extends to $k$-forms (i.e., rank-$k$, anti-symmetric covariant tensors) using the wedge product. In particular, this means that the bundle of $k$-forms on $M$ decomposes as:
\be\label{kformdecomp}
\Omega^{k}(M)_{\C}=\bigoplus_{p+q=k} \Omega^{p,q}(M)\,,
\ee
where a section of $\Omega^{p,q}(M)$ has $p$ holomorphic form indices and $q$ anti-holomorphic form indices:
\begin{equation*}
 \omega\in\Omega^{p,q}(M)\,, \quad \omega= \omega_{a_1 \cdots a_p \bar{a}_{1}\cdots\bar{a}_{q}}\,\d z^{a_1} \wedge \cdots\wedge\d z^{a_p}\wedge\d\bar{z}^{\bar{a}_1}\wedge\cdots\wedge\d\bar{z}^{\bar{a}_q}\,.
\end{equation*}
Clearly, $\Omega^{p,q}(M)=\emptyset$ whenever $p+q>2d$ or $p,q>d$, where $d$ is the complex dimension of $M$.

Let $\rho_{p,q}:\Omega^{k}(M)_{\C}\rightarrow\Omega^{p,q}(M)$ be the natural projection onto $(p,q)$-forms. We can now define a \emph{Dolbeault operator}, $\dbar$, which increases the anti-holomorphic form degree of any tensor by one:
\be\label{dol1}
\dbar:\Omega^{p,q}(M)\rightarrow \Omega^{p,q+1}(M)\,, \qquad \dbar|_{\Omega^{p,q}(M)}=\rho_{p,q+1}\circ \d\,,
\ee
where $\d$ is the usual exterior derivative. We take the differential operator $\dbar$ to be our working definition of an almost complex structure. Indeed, this coincides with our intuitive definition: $\dbar$ is precisely the operator which distinguishes between holomorphic and anti-holomorphic degrees of freedom. For instance, given any function $f$ on $M$, the condition that $f$ be holomorphic is simply $\dbar f=0$. In this language, an almost complex structure $\dbar$ is \emph{integrable} if $\dbar^2=0$; by the Newlander-Nirenberg theorem, this is equivalent to the vanishing Nijenhuis tensor condition for the underlying $J$ given above.

\medskip

Twistor space is an open subset of $\CP^3$, which is naturally a complex manifold (of complex dimension $3$ or real dimension $6$). So given a notion of complex conjugation, it is clear that the complex structure on $\PT$ is given by
\be\label{PTcs1}
\dbar=\d\bar{Z}^{\bar{A}}\frac{\partial}{\partial\bar{Z}^{\bar{A}}}\,.
\ee
We have seen that what exactly we mean by the complex conjugation here depends on what sort of real signature slice of $\M_\C$ we want to describe. Since it will be our preferred choice of reality structure in subsequent lectures, we can explicitly write down this complex structure in the Euclidean reality conditions.

Since the twistor space of $\R^4$ is the projective spinor bundle, there are natural bases for the anti-holomorphic vectors and the $(0,1)$-forms on $\PT$:
\be\label{Tbar}
T_{\PT}^{0,1}=\mathrm{span}\left\{\dbar_0=\la\lambda\,\hat{\lambda}\ra \lambda^{\alpha}\frac{\partial}{\partial\hat{\lambda}^{\alpha}},\,\dbar_{\dot\alpha}=\lambda^{\alpha}\frac{\partial}{\partial x^{\alpha\dot\alpha}}\right\}\,,
\ee
\be\label{T*bar}
\Omega^{0,1}(\PT)=\mathrm{span}\left\{\bar{e}^{0}=\frac{\la\hat{\lambda}\,\d\hat{\lambda}\ra}{\la\lambda\,\hat{\lambda}\ra^2},\,\bar{e}^{\dot\alpha}=\frac{\hat{\lambda}_{\alpha}\,\d x^{\alpha\dot\alpha}}{\la\lambda\,\hat{\lambda}\ra}\right\}\,.
\ee
With these bases, the complex structure on twistor space is given by:
\be\label{EuCS}
\dbar=\bar{e}^{0}\,\dbar_{0}+\bar{e}^{\dot\alpha}\,\dbar_{\dot\alpha}\,.
\ee
It is easy to see that this is compatible with the twistor correspondence, in the sense that
\begin{equation*}
 \bar{e}^{0}\,\dbar_{0}+\bar{e}^{\dot\alpha}\,\dbar_{\dot\alpha}=\d\hat{\mu}^{\dot\alpha}\frac{\partial}{\partial\hat{\mu}^{\dot\alpha}}+\d\hat{\lambda}^{\alpha}\frac{\partial}{\partial\hat{\lambda}_{\alpha}}=\d\hat{Z}^{A}\,\frac{\partial}{\partial\hat{Z}^{A}}\,.
\end{equation*}
This follows straightforwardly from the incidence relations. Furthermore, you can easily convince yourself that this is an integrable complex structure: $\dbar^2=0$.

\subsection{Conformal structures}

Thus far, we have been very na\"ive regarding the conformal structure of space-time. The null cones associated with points in $\M_\C$ or its real slices are encoded in twistor space by the intersections of the corresponding twistor lines. We happily stated that this amounts to capturing everything about the conformal structure of $\M_\C$ (or its real slices) in terms of holomorphic structures on $\PT$. But light cones are not everything: these only capture the conformal structure of space-time up to boundary conditions. In other words, knowing about light cones is only enough to identify the conformal class of a space-time: in this, case, the class of conformally flat metrics. 

In standard language, we can make the distinction between Minkowski space and other conformally flat spaces (e.g., dS$_4$) by saying what the space-time looks like `at infinity.' This can be made precise using Penrose's notion of conformal compactification. The conformal infinity $\scri$ of Minkowski space has the structure of three points (space-like infinity $i^0$ and future/past time-like infinity $i^{\pm}$) and the null hypersurfaces $\scri^{\pm}$ of topology $\R\times S^2$ joining them. By contrast, the conformal infinity of dS$_4$ is composed of two space-like three-spheres which form the past and future time-like infinities.

It is easy to see that, as it stands, twistor space is not sensitive to the different conformal structures within the class of conformally flat space-times. The complexified conformal group in four-dimensions is $\SL(4,\C)$, and it is easy to see that we can form $\SL(4,\C)$-invariants from any four distinct points in $\PT$ using only the four-dimensional Levi-Civita symbol, $\epsilon_{ABCD}$:
\be\label{ConInv1}
(Z_1,Z_2,Z_3,Z_4):=\epsilon_{ABCD}\,Z_1^{A}\,Z_2^{B}\,Z_3^{C}\,Z_4^{D}\,.
\ee
More generally, twistor space carries a natural un-broken action of the complexified conformal group. One way of seeing this is to show that twistor indices are actually spinor indices of $\SL(4,\C)$, but we can also just construct a representation of $\SL(4,\C)$ which acts on $\PT$ explicitly.

Such a representation will have generators $T^{A}_{B}$, acting as $Z^{A}\rightarrow T^{A}_{B}Z^{B}$. Crucially, we can find a representation of $\SL(4,\C)$ for which these generators are \emph{linear}:
\be\label{ConGen1}
T^{A}_{B}=Z^{A}\,\frac{\partial}{\partial Z^{B}}\,,
\ee
for different values of the twistor indices. Note that these generators are holomorphic, as they must be, since we already know that the causal structure (i.e., light cones) of conformally flat spaces are captured by the holomorphic structure on twistor space.

In standard notation, the generators of the conformal group are written in twistor space as:
\be\label{ConGen2}
P_{\alpha\dot\alpha}=\lambda_{\alpha}\,\frac{\partial}{\partial\mu^{\dot\alpha}}\,, \qquad J_{\alpha\beta}=\lambda_{(\alpha}\,\frac{\partial}{\partial\lambda^{\beta)}}\,, \qquad \tilde{J}_{\dot\alpha\dot\beta}=\mu_{(\dot\alpha}\,\frac{\partial}{\partial\mu^{\dot\beta)}}\,,
\ee
\begin{equation*}
 K^{\alpha\dot\alpha}=\mu^{\dot\alpha}\,\frac{\partial}{\partial\lambda_{\alpha}}\,, \qquad D=\frac{1}{2}\left(\lambda_{\alpha}\frac{\partial}{\partial\lambda_{\alpha}}-\mu^{\dot\alpha}\frac{\partial}{\partial\mu^{\dot\alpha}}\right)\,,
\end{equation*}
with the identifications of $P_{\alpha\dot\alpha}$, $J_{\alpha\beta}$, $\tilde{J}_{\dot\alpha\dot\beta}$ as the generators of Lorentz boosts and rotations, $K^{\alpha\dot\alpha}$ the generator of special conformal transformations, and $D$ the dilatation generator. You may find it an interesting exercise to confirm for yourself that the commutators of these operators in twistor space do indeed generate the conformal algebra.

The fact that $\PT$ carries a linear action of the conformal group means that there is no way for us to distinguish between conformally flat space-times. In particular, if we really want the twistor space of $\M_\C$, some additional structure is required on $\PT$ which breaks conformal invariance. By comparison with the space-time perspective, it's clear that this missing structure must have something to do with the `points at infinity' associated with specifying the conformal structure. So how do we determine the structure of `infinity' on twistor space?

\medskip

Any conformally flat metric can be written as
\be\label{Confl1}
\d s^{2}=\frac{\d x^{\alpha\dot\alpha}\,\d x_{\alpha\dot\alpha}}{(f(x))^2}\,,
\ee
for some function $f(x)$, which is the conformal factor relating the metric to the flat (Minkowski) one. \emph{A priori}, twistor space can't tell the difference between this metric and the true Minkowski one, for which $f(x)=1$. To see what structure is needed on twistor space to differentiate between conformally flat metrics, we can try to write the metric \eqref{Confl1} in terms of twistor space quantities.

Recall that a point in conformally flat space-time is represented by a line in twistor space, and you showed that these lines are in turn represented by skew bi-twistors
\be\label{Confl2}
X^{AB}=\la\lambda_{1}\,\lambda_{2}\ra\,  \left(\begin{array}{c c}
                                                                            \frac{1}{2}\epsilon^{\dot{\alpha}\dot{\beta}} x^{2} & x^{\dot\alpha}_{\beta} \\
                                                                            -x^{\dot\beta}_{\alpha} & \epsilon_{\alpha\beta}
                                                                            \end{array}\right)\,,
\ee
where $Z_{1}^{A}$, $Z_{2}^{B}$ are any two points lying on the line $X\cong\CP^1$. There is a natural line element we can write in terms of the bi-twistor variables:
\be\label{Confl3}
\d s^{2}=\epsilon_{ABCD}\,\d X^{AB}\,\d X^{CD}\,.
\ee
This metric is obviously flat (since the metric components $\epsilon_{ABCD}$ are constants); is it in fact the Minkowski metric? The answer is no: $X^{AB}$ encodes a point in space-time up to a scale, corresponding to $\la\lambda_{1}\lambda_{2}\ra$ in \eqref{Confl2}. This means that if we want to interpret the $X^{AB}$ as space-time coordinates, then we must consider them only up to scale; in other words, we must treat them as homogeneous coordinates.\footnote{A general skew-symmetric $X^{AB}$ contains six degrees of freedom. Quotienting by projective rescalings means that the $X^{AB}$ can be treated as homogeneous coordinates on $\CP^5$, reducing the degrees of freedom to five. The fact that $X^{AB}$ is formed from the skew of two vectors (i.e., $Z_1$ and $Z_2$) is equivalent to saying that $X^2=\epsilon_{ABCD}X^{AB} X^{CD}=0$. So the quadric $Q=\{X\in\CP^5|X^2=0\}$ has four degrees of freedom. This is something known as the \emph{Klein quadric}, which represents points in $Q$ as lines in a complex projective 3-space, namely, twistor space.} Clearly, the line element \eqref{Confl3} is not homogeneous of degree zero, so it is not projectively well-defined.

Indeed, working with \eqref{Confl2} you can show that the line element \eqref{Confl3} is
\be\label{Confl4}
\d s^{2}=\la\lambda_{1}\,\lambda_{2}\ra^{2}\,\d x^{\alpha\dot\alpha}\,\d x_{\alpha\dot\alpha}\,,
\ee
which is the Minkowski metric up to a scale. Thus, \eqref{Confl3} is the form of the conformally flat metric, written in terms of the skew bi-twistor coordinates for space-time points. In order to get a metric in a particular conformal structure, we have to write the line element in a projectively invariant fashion. Since \eqref{Confl3} has homogeneous weight $+2$, such a line element will take the form:
\be\label{Confl5}
\d s^{2}= \frac{\epsilon_{ABCD}\,\d X^{AB}\,\d X^{CD}}{(I_{AB}\,X^{AB})^2}\,,
\ee
for some fixed skew bi-twistor $I_{AB}$. This metric is singular on the hypersurface $I_{AB}X^{AB}=0$, which defines a set of points `at infinity' in the usual sense of conformal compactification. 

So $I_{AB}$ is the ingredient required to break conformal invariance on twistor space. It encodes the structure of the hypersurface at infinity in space-time and thus the conformal structure. For this reason, it is known as the \emph{infinity twistor}. This infinity twistor is precisely the reason why twistor space is required to be an \emph{open subset} of $\CP^3$, rather than the entire projective space itself. If the lines $X\subset\CP^3$ for which $I_{AB}X^{AB}=0$ correspond to points which lie at infinity in space-time, then clearly such lines should not be included in $\PT$. In other words, $\PT$ should be the open subset of $\CP^3$ for which all lines contained in $\PT$ satisfy $I_{AB}X^{AB}\neq 0$. In other words, which open subset of $\CP^3$ we take to be $\PT$ depends upon which conformal structure we choose for space-time.

\medskip

We started out with the goal of representing the flat conformal structure of $\M_\C$ in twistor space. We've now established that this requires an appropriate choice of infinity twistor. Consider the choice
\be\label{inft1}
I_{AB}=\frac{1}{2}\,\left(\begin{array}{c c}
                           0 & 0 \\
                           0 & \epsilon^{\alpha\beta}
                          \end{array}\right)\,.
\ee
It is easy to see that
\begin{equation*}
 I_{AB}\,X^{AB}=\frac{\la\lambda_{1}\lambda_{2}\ra}{2}\,\epsilon^{\alpha\beta}\,\epsilon_{\alpha\beta}=\la\lambda_{1}\,\lambda_{2}\ra\,,
\end{equation*}
so the line element \eqref{Confl5} with this infinity twistor is indeed the complexified Minkowski metric. 

This infinity twistor also makes sense from a twistor space point of view. Consider a line in $\PT$ for which $I_{AB} X^{AB}=0$; since $I_{AB} X^{AB}=\la\lambda_{1}\lambda_{2}\ra$, this means that $\lambda_{1\,\alpha}\propto\lambda_{2\,\alpha}$. But since both points $Z_{1},Z_{2}$ lie on the same line $X$ in twistor space, the only way that their undotted spinor components can be proportional is if they are both zero. So $I_{AB}X^{AB}=0$ for the infinity twistor \eqref{inft1} if and only if the points lying on $X$ have the form $Z^{A}=(\mu^{\dot\alpha},0)$. On the other hand, these points should obey the incidence relations $\mu^{\dot\alpha}=x^{\alpha\dot\alpha}\lambda_{\alpha}$. If $\lambda_{\alpha}=0$ and $x^{\alpha\dot\alpha}$ is finite, then $\mu^{\dot\alpha}=0$ as well. However, $Z^{A}$ are homogeneous coordinates, which means that we cannot have $\mu^{\dot\alpha}=0$ and $\lambda_{\alpha}=0$ simultaneously. This means that some component of $x^{\alpha\dot\alpha}$ must be infinitely large if $\lambda_\alpha=0$. This is exactly what we expect: lines in $\PT$ for which $I_{AB}X^{AB}=0$ should correspond to points at infinity in $\M_\C$!

Furthermore, it is easy to see that $\epsilon^{ABCD}I_{AB}I_{CD}=0$, which means that the infinity twistor \eqref{inft1} corresponds to a line $I$ in $\PT$. This line in twistor space is precisely the space-like infinity of Minkowski space, which is a point $i^0$ in the conformal compactification. Lines in twistor space which intersect $I$ correspond to points in space-time which are null separated from $i^0$; these are the points of null infinity, $\scri^{\pm}$. So the infinity twistor really does encode all the information associated with the conformal structure of space-time.

\subsubsection*{Exercise: \textit{the twistor space of Euclidean} AdS$_4$}

This exercise involves applying both reality and conformal structures to write down the twistor space of another conformally flat space: Euclidean AdS$_4$. In standard Cartesian coordinates on the general conformally flat space-time, let $x^0=r$ be the radial direction of Poincar\'e coordinates. First, determine how to write the Poincar\'e metric on Euclidean AdS$_4$ in the spinor formalism (you'll need to impose some reality conditions on $x^{\alpha\dot\alpha}$, and it might be useful to write down the spinor form of the unit normal to the boundary). Next, find the infinity twistor appropriate to Euclidean AdS$_4$ -- what $I_{AB}$ is required in \eqref{Confl5} to obtain the metric that you just wrote down? Finally, what is the corresponding twistor space?

In then end, you should find that the twistor space of Euclidean AdS$_4$ is
\be\label{Ads4}
\PT^{+}=\left\{Z\in\PT | Z\cdot\bar{Z}> 0\right\}\,.
\ee
Surprisingly, the Lorentzian notion of complex conjugate (i.e., $\bar{Z}_{A}$) enters here, despite the fact that you are describing a Euclidean space-time. If you're having trouble seeing why, remember that the AdS-boundary is defined by $I_{AB} X^{AB}=0$, for the infinity twistor you wrote down. Think about how you write a Euclidean real $X^{AB}$ in twistor space, and how this expression contracts with the infinity twistor.

%% file: TLecture3.tex
\section{The Penrose Transform}

Now that we have explored the basic geometry of twistor theory, it is natural to ask: what is it good for? In this lecture we will explore one of the oldest applications of twistor theory: providing solutions to free field equations. As we will see, every massless free field of integer or half-integer spin in four-dimensional flat space-time can be represented on twistor space by a piece of geometric data called a \emph{cohomology class} -- a differential form which obeys some simple differential equations.


\subsection{Zero-rest-mass fields}

In physics, we often deal with free fields. For instance, if we want to compute a scattering amplitude in some quantum field theory, the asymptotic states in the scattering process are taken to be free fields; the LSZ reduction formula imposes the free equations of motion on the external states. We usually think of such free fields in terms of gauge potentials. Let's focus on the case of massless free fields; for spin zero this is just a massless scalar $\Phi$; for spin one we have the Maxwell field $A_{a}$, for spin two the linearized metric $h_{ab}$, and so on. 

Of course, for integer spins greater than zero this is not an invariant way of thinking about free fields: different potentials can describe the same physical field if they differ by gauge transformations. For the Maxwell field, these are the usual transformations $A_{a}\rightarrow A_{a}+\partial_{a}\lambda$, while for the metric these are linearized diffeomorphisms. The familiar objects which are invariant under gauge transformations are the linearized curvature tensors associated with the spin-$s$ fields. In four-dimensions, certain underlying structures of these invariant objects become manifest when working in the 2-spinor formalism. This enables us to write the free field equations for massless spin-$s$ fields in terms of these underlying structures.

To illustrate how this works, let's start with spin one. The usual 2-spinor yoga tells us that the Maxwell gauge potential $A_{a}$ can be translated into an object with two spinor indices, $A_{\alpha\dot\alpha}$. Its field strength is therefore
\be\label{MaxFS1}
F_{\alpha\dot\alpha\beta\dot\beta}=\partial_{\alpha\dot\alpha}A_{\beta\dot\beta}-\partial_{\beta\dot\beta}A_{\alpha\dot\alpha}\,.
\ee
By definition, this field strength is anti-symmetric under the exchange of $(\alpha\dot\alpha)\leftrightarrow(\beta\dot\beta)$; this is just the spinor version of the usual anti-symmetry $F_{ab}=-F_{ba}$. Clearly, there are only two ways that such an anti-symmetry can arise: either a contribution to $F$ is skew symmetric in $\alpha\leftrightarrow\beta$ and symmetric under $\dot\alpha\leftrightarrow\dot\beta$, or it must be the other way around. Anything which is skew in two un-dotted spinor indices must be proportional to $\epsilon_{\alpha\beta}$, and similarly for dotted spinor indices. So we can write this decomposition as
\be\label{MaxFS2}
F_{\alpha\dot\alpha\beta\dot\beta}=\frac{1}{2}\,\epsilon_{\alpha\beta}\,F^{\gamma}{}_{\dot\alpha\gamma\dot\beta}+\frac{1}{2}\,\epsilon_{\dot\alpha\dot\beta}\,F_{\alpha}{}^{\dot\gamma}{}_{\beta\dot\gamma}\,.
\ee
It's easy to see that the contracted pieces of $F$ appearing in this expression are symmetric in their remaining free spinor indices, so we can define the quantities
\be\label{MaxFS3}
\tilde{F}_{\dot\alpha\dot\beta}=\tilde{F}_{(\dot\alpha\dot\beta)}:=\frac{1}{2}\,F^{\gamma}{}_{\dot\alpha\gamma\dot\beta}\,, \qquad F_{\alpha\beta}=F_{(\alpha\beta)}:=\frac{1}{2}F_{\alpha}{}^{\dot\gamma}{}_{\beta\dot\gamma}\,,
\ee
which will be referred to as the \emph{self-dual} (SD) and \emph{anti-self-dual} (ASD) portions of the field strength, respectively.

With this new notation, the field strength is
\be\label{MaxFS4}
F_{\alpha\dot\alpha\beta\dot\beta}=\epsilon_{\alpha\beta}\,\tilde{F}_{\dot\alpha\dot\beta}+\epsilon_{\dot\alpha\dot\beta}\,F_{\alpha\beta}\,.
\ee
It is easy to see why we have chosen the names `self-dual' and `anti-self-dual' for the two non-trivial portions of the field strength. Recall that we can always form the dual field strength in standard notation by contracting with the 4-dimensional Levi-Civita symbol: $\epsilon^{abcd} F_{ab}$. In Euclidean signature, the Levi-Civita symbol is translated into 2-spinors as:
\be\label{4dLC}
\epsilon^{abcd}\leftrightarrow \epsilon^{\alpha\gamma}\,\epsilon^{\beta\delta}\,\epsilon^{\dot\alpha\dot\delta}\,\epsilon^{\dot\beta\dot\gamma}-\epsilon^{\alpha\delta}\,\epsilon^{\beta\gamma}\,\epsilon^{\dot\alpha\dot\gamma}\,\epsilon^{\dot\beta\dot\delta}\,,
\ee
and a straightforward calculation shows that
\be\label{dualFS}
\frac{1}{2}\,\epsilon^{abcd}\,F_{ab}=\epsilon^{\gamma\delta}\,\tilde{F}^{\dot\gamma\dot\delta}-\epsilon^{\dot\gamma\dot\delta}\,F^{\gamma\delta}\,.
\ee
So $\tilde{F}_{\dot\alpha\dot\beta}$ has eigenvalue $+1$ under the duality operation, while $F_{\alpha\beta}$ has eigenvalue $-1$.

Written in terms of the SD/ASD decomposition, the Maxwell equations and Bianchi identity for the field strength are
\be\label{Maxeq}
\partial^{\dot\alpha}_{\beta} \tilde{F}_{\dot\alpha\dot\beta}+\partial^{\alpha}_{\dot\beta}F_{\alpha\beta}=0\,,
\ee
\be\label{Bianchi}
\partial^{\dot\alpha}_{\beta} \tilde{F}_{\dot\alpha\dot\beta}-\partial^{\alpha}_{\dot\beta}F_{\alpha\beta}=0\,,
\ee
respectively. Recall that the Bianchi identity is non-dynamical: any field strength obeys \eqref{Bianchi}. These two equations allow us to see that purely SD or ASD Maxwell fields are consistent solutions to the equations of motion. Indeed, a purely SD gauge field is characterized by $F_{\alpha\beta}=0$. With this constraint, the remaining components of the Maxwell equation and Bianchi identity are equivalent:
\be\label{SDgf}
\partial^{\dot\alpha}_{\beta}\,\tilde{F}_{\dot\alpha\dot\beta}=0\,,
\ee
so this equation is automatically satisfied. A similar argument works for the purely ASD sector, $\tilde{F}_{\dot\alpha\dot\beta}=0$.

This means that the SD and ASD parts of the field strength can be considered separately, each defining a consistent on-shell sector. These are precisely the two on-shell photon polarizations we expect in four-dimensions, often referred to as positive or negative \emph{helicity}. A Maxwell field which is purely SD (i.e., $F_{\alpha\beta}=0$) is identified with the positive helicity polarization, while a purely ASD (i.e., $\tilde{F}_{\dot\alpha\dot\beta}=0$) field is identified with the negative helicity polarization. 

Working with this SD/ASD (or positive/negative helicity) decomposition of the field strength means that we can phrase the free-field equations of motion purely in terms of the field strength components. Given some symmetric $\tilde{F}_{\dot\alpha\dot\beta}$, what is the condition for this to describe a SD (positive helicity) Maxwell field? The answer is provided by \eqref{SDgf}:
\be\label{Maxzrm1}
\partial^{\dot\alpha}_{\beta}\,\tilde{F}_{\dot\alpha\dot\beta}=0\,.
\ee
Similarly, a symmetric $F_{\alpha\beta}$ describes a ASD (negative helicity) Maxwell field provided that
\be\label{Maxzrm2}
\partial^{\alpha}_{\dot\beta}\,F_{\alpha\beta}=0\,.
\ee
These equations are the spin-1 \emph{zero-rest-mass} (z.r.m.) equations: they constitute the free field equations for Maxwell fields, formulated in terms of the SD/ASD components of the field strength.

\medskip

A similar story holds for any integer or half-integer spin: the (gauge-invariant) curvature tensor associated to the spin-$s$ gauge field contains SD and ASD components which define consistent subsectors of the equations of motion. For example, the Riemann curvature tensor (corresponding to $s=2$) can be decomposed as
\begin{multline}\label{Riemann}
 R_{abcd}\leftrightarrow \epsilon_{\alpha\beta}\,\epsilon_{\gamma\delta}\,\widetilde{\Psi}_{\dot\alpha\dot\beta\dot\gamma\dot\delta} + \epsilon_{\alpha\beta}\,\epsilon_{\gamma\delta}\,\Psi_{\alpha\beta\gamma\delta} +\epsilon_{\dot\alpha\dot\beta}\,\epsilon_{\gamma\delta}\,\Phi_{\alpha\beta\dot\gamma\dot\delta}
 +\epsilon_{\alpha\beta}\,\epsilon_{\dot\gamma\dot\delta}\,\Phi_{\gamma\delta\dot\alpha\dot\beta} \\
 +\frac{R}{12}\left(\epsilon_{\alpha\gamma}\,\epsilon_{\beta\delta}\,\epsilon_{\dot\alpha\dot\beta}\,\epsilon_{\dot\gamma\dot\delta}+\epsilon_{\alpha\beta}\,\epsilon_{\gamma\delta}\,\epsilon_{\dot\alpha\dot\gamma}\,\epsilon_{\dot\beta\dot\delta}\right)\,
\end{multline}
with $\widetilde{\Psi}_{\dot\alpha\dot\beta\dot\gamma\dot\delta}$ and $\Psi_{\alpha\beta\gamma\delta}$ totally symmetric, encoding the SD and ASD portions of the Weyl curvature; $\Phi_{\alpha\beta\dot\gamma\dot\delta}$ encoding the trace-free Ricci curvature; and $R$ the Ricci scalar. The vacuum Einstein equations enforce $\Phi_{\alpha\beta}^{\dot\gamma\dot\delta}=0=R$; on the support of these equations the Bianchi identity $\nabla_{[a}R_{bc]de}=0$ is equivalent to
\be\label{gBianchi}
\epsilon_{\gamma\delta}\,\nabla^{\dot\alpha}_{\beta}\,\widetilde{\Psi}_{\dot\alpha\dot\beta\dot\gamma\dot\delta}-\epsilon_{\dot\gamma\dot\delta}\,\nabla^{\alpha}_{\dot\beta}\,\Psi_{\alpha\beta\gamma\delta}=0\,.
\ee
From this we see that the SD (i.e., $\Psi=0$) and ASD (i.e., $\widetilde{\Psi}=0$) sectors are consistent, subject to the Bianchi identities
\be\label{gBianchi2}
\nabla^{\dot\alpha}_{\beta}\,\widetilde{\Psi}_{\dot\alpha\dot\beta\dot\gamma\dot\delta}=0\,, \qquad \nabla^{\alpha}_{\dot\beta}\,\Psi_{\alpha\beta\gamma\delta}=0\,,
\ee
respectively. Linearizing these equations by replacing the covariant derivatives with partial derivatives gives the helicity $\pm2$ z.r.m. equations.

In general, a z.r.m. field of helicity $h$ (for $h$ any integer or half-integer) is represented by a field with $2|h|$ dotted or un-dotted symmetric spinor indices (depending upon the sign of $h$) which obeys a linear PDE:
\begin{equation*}
 h>0 \quad \tilde{\phi}_{\dot\alpha_1\cdots\dot\alpha_{2|h|}}\,, \qquad \partial^{\beta\dot\alpha_{1}}\,\tilde{\phi}_{\dot\alpha_1\cdots\dot\alpha_{2|h|}}=0\,, 
\end{equation*}
\be\label{zrms}
h=0 \quad \Phi\,, \qquad \Box\Phi=\partial^{\alpha\dot\alpha}\partial_{\alpha\dot\alpha}\,\Phi=0\,,
\ee
\begin{equation*}
 h<0 \quad \phi_{\alpha_{1}\cdots\alpha_{2|h|}}\,, \qquad \partial^{\alpha_{1}\dot\beta}\,\phi_{\alpha_{1}\cdots\alpha_{2|h|}}=0\,.
\end{equation*}
As desired, this gives a representation of free fields in terms of their linearized SD or ASD field strengths. From now on, when we refer to free fields of a given helicity, we will implicitly have in mind this z.r.m. field representation.

\medskip

This representation associates two totally symmetric spinors (one dotted, one un-dotted) with any field of spin $s>0$; these spinors encode the information contained in the totally trace-free portion of the linearized spin $s$ curvature tensor associated with the field. For the spin-1 case, this is the entire field strength, while for spin-2 it's the Weyl tensor. This general splitting of a trace-free curvature tensor into SD and ASD parts is a special feature of four-dimensions. You may have heard a more sophisticated geometric explanation for this splitting before, so it's worth mentioning it here.

Trace-free curvature tensors can always be represented as 2-forms on space-time: this was obvious in the Maxwell field case we covered above. On any 4-dimensional Riemannian manifold $M^4$, the space of 2-forms has a special property: it is preserved by the Hodge star (in coordinates, this is just the duality operator defined by $\epsilon^{abcd}$), which acts involutively:
\be\label{Hodgestar}
*:\Omega^{2}(M^4)\rightarrow\Omega^{2}(M^4)\,, \qquad *^2=\mathrm{id}\,.
\ee
This means that any 2-form can be decomposed into components which have eigenvalue $\pm1$ with respect to the Hodge star,
\be\label{Hodgestar2}
\Omega^{2}(M^4)=\Omega^{2}_{+}(M^4)\oplus\Omega^{2}_{-}(M^4)\,.
\ee
This decomposition is precisely the decomposition into SD and ASD parts that we worked out in spinor components above! This is yet another advantage of the 2-spinor formalism: it allows us to manifest the decomposition \eqref{Hodgestar2} in terms of totally symmetric spinors of different chirality.

\subsubsection*{\textit{Conformal invariance}}

Besides allowing us to work directly with gauge-invariant representations of free fields, the z.r.m. equations have another interesting property: they are conformally invariant. To see this, consider a conformal re-scaling of $\M_{\C}$,
\be\label{contrans1}
\eta_{ab}\rightarrow \Omega^{2}(x)\,\eta_{ab}\,.
\ee
In the 2-spinor language, the complexified metric is represented by $\eta_{ab}\leftrightarrow\epsilon_{\alpha\beta}\epsilon_{\dot\alpha\dot\beta}$, so it is natural to declare that each factor transforms with the same weight:
\be\label{contrans2}
\epsilon_{\alpha\beta}\rightarrow \Omega(x)\,\epsilon_{\alpha\beta}\,, \qquad \epsilon_{\dot\alpha\dot\beta}\rightarrow\Omega(x)\,\epsilon_{\dot\alpha\dot\beta}\,.
\ee
Under such a conformal transformation, it turns out that all z.r.m. fields transform with a factor of $\Omega^{-1}$. It is easy to convince yourself why this is true for $s=1$ (just use the decomposition \eqref{MaxFS4} and fact that $F_{ab}$ is conformally-invariant), and similar arguments work for any other spin. 

For concreteness, consider the negative helicity z.r.m. equation
\be\label{contrans3}
\partial^{\alpha\dot\alpha}\phi_{\alpha\beta\cdots\gamma}=0.
\ee
Using the definitions \eqref{contrans2}, the fact that $\phi_{\alpha\cdots\beta}$ has conformal weight $-1$, and $\partial_{\alpha\dot\alpha}\epsilon_{\beta\gamma}=0$, you can deduce that
\begin{multline}\label{contrans4}
 \Omega\,\hat{\nabla}_{\alpha\dot\alpha}\hat{\phi}_{\beta\cdots\gamma}=\Omega\,\hat{\nabla}_{\alpha\dot\alpha}\left(\Omega^{-1}\,\phi_{\beta\cdots\gamma}\right) \\
 = \partial_{\alpha\dot\alpha}\phi_{\beta\cdots\gamma}-\Upsilon_{\dot\alpha\alpha}\,\phi_{\beta\cdots\gamma}-\Upsilon_{\dot\alpha\beta}\,\phi_{\alpha\cdots\gamma}-\cdots-\Upsilon_{\dot\alpha\gamma}\,\phi_{\beta\cdots\alpha}\,,
\end{multline}
where hatted quantities indicate objects in the conformally re-scaled metric, and
\be\label{upsilon}
\Upsilon_{\dot\alpha\alpha}:=\frac{1}{k}\,\Omega^{-k}\,\partial_{\alpha\dot\alpha}\Omega^{k}\,, \qquad \forall k\in\Z\,.
\ee
Contracting both sides of \eqref{contrans4} with $\Omega^{-2}\epsilon^{\alpha\beta}\epsilon^{\dot\alpha\dot\beta}$ yields
\be\label{contrans5}
\hat{\nabla}^{\beta\dot\beta}\hat{\phi}_{\beta\cdots\gamma}=\Omega^{-3}\,\partial^{\beta\dot\beta}\phi_{\beta\cdots\gamma}\,.
\ee
Thus, if the z.r.m. equation \eqref{contrans3} is satisfied in Minkowski space-time, then it will also be satisfied in any conformally flat space-time. A similar argument works for the positive helicity z.r.m. equations. In the scalar case, it follows that the massless scalar obeys the \emph{conformally-coupled} wave equation in the conformally flat space-time:
\be\label{concouple}
\left(\Box+\frac{R}{6}\right)\Phi=0\,,
\ee
for $R$ the scalar curvature of the conformally re-scaled metric.


\subsection{The Penrose transform}

The z.r.m. equations are a conformally-invariant way of encoding the free field equations. In the previous lecture, we established that conformal invariance is naturally encoded in twistor space (and only broken by the choice of an additional structure -- the infinity twistor). A natural question is then: can we use twistor theory to generate solutions to the z.r.m. equations?

Consider a negative helicity solution to the spin $s$ z.r.m. equations; this is a totally symmetric spinor field $\phi_{\alpha_1\cdots\alpha_{2s}}(x)$ on $\M_\C$ which obeys
\be\label{nhPT1}
\partial^{\alpha_{1}\dot\alpha}\,\phi_{\alpha_1\cdots\alpha_{2s}}=0\,.
\ee
Clearly, such a field is local on space-time, and we know that a point $x\in\M_{\C}$ corresponds to a line $X\cong\CP^1$ inside twistor space. So if we want to find a twistorial way of encoding the field $\phi_{\alpha_1\cdots\alpha_{2s}}$, the $\CP^1$ degrees of freedom on twistor space must be removed in some way. One way of doing this is to integrate them out explicitly.

Furthermore, we need to build an object which has $2s$ symmetric, un-dotted spinor indices. This suggests some sort of twistor space construction of the form:
\be\label{nhPT2}
\phi_{\alpha_1\cdots\alpha_{2s}}(x)\stackrel{?}{=} \int\limits_{X\cong\CP^1}\la\lambda\,\d\lambda\ra\, \lambda_{\alpha_1}\cdots\lambda_{\alpha_{2s}}\,(\cdots)\,,
\ee
where $\la\lambda\,\d\lambda\ra$ is the natural holomorphic measure on $\CP^1$ of projective weight $+2$ and the $(\cdots)$ stands for some other ingredients which are yet to be determined. The form of these extra ingredients is tightly constrained simply by requiring that the integral is well-defined.

For \eqref{nhPT2} to make sense, the integrand must be a $(1,1)$-form on $X$ of homogeneity zero. Excluding the missing ingredients, the portion of the integrand we have written out so far is a $(1,0)$-form of homogeneity $2s+2$. Thus, we must have $(\cdots)=f(\lambda, \bar{\lambda})$, where $f$ is a weight $-2s-2$ $(0,1)$-form on $\CP^1$. In other words, 
\begin{equation*}
f(\lambda,\bar{\lambda})=f^{\bar{\alpha}}(\lambda,\bar{\lambda})\,\d\bar{\lambda}_{\bar{\alpha}}\,, \qquad f(r\lambda,\,\bar{r}\bar{\lambda})=r^{-2s-2}\,f(\lambda,\bar{\lambda})\,.
\end{equation*}
Such an object is naturally provided by a $(0,1)$-form on twistor space of homogeneity $-2s-2$ which we restrict to $X\cong \CP^1$ using the incidence relations. We denote such an object as
\be\label{difform}
f\in\Omega^{0,1}(\PT,\cO(-2s-2))\,,
\ee
which should be read as: `$f$ is a $(0,1)$-form on $\PT$ of projective weight $-2s-2$.' The restriction to $X$ is implemented by
\be\label{Xrestrict}
f(Z,\bar{Z})|_{X}=f(x^{\beta\dot\alpha}\lambda_{\beta},\lambda_{\alpha},\,\overline{x^{\beta\dot\alpha}\lambda_{\beta}}, \bar{\lambda}_{\bar{\alpha}})\,,
\ee
leaving us with precisely the sort of object we need to complete \eqref{nhPT2}.

Putting all of the ingredients together, we are left with a proposal for the negative helicity, spin $s$ z.r.m. field of the form:
\be\label{nhPT3}
 \phi_{\alpha_1\cdots\alpha_{2s}}(x)=\int_{X}\la\lambda\,\d\lambda\ra\wedge \lambda_{\alpha_1}\cdots\lambda_{\alpha_{2s}}\,f(Z)|_{X}\,.
\ee
This results in a well-defined space-time field of the appropriate helicity, but it's not at all clear that this field satisfies the z.r.m. equation \eqref{nhPT1}. To check this, we simply compute
\be\label{nhPT4}
 \partial^{\alpha_{1}\dot\alpha}\,\phi_{\alpha_1\cdots\alpha_{2s}}=\int_{X}\la\lambda\,\d\lambda\ra\wedge\lambda_{\alpha_1}\cdots\lambda_{\alpha_{2s}}\left(\lambda^{\alpha_1}\,\left.\frac{\partial f}{\partial\mu_{\dot\alpha}}\right|_{X} \right.\\
 \left.+\overline{\lambda^{\alpha_1}}\,\left.\frac{\partial f}{\partial\overline{\mu_{\dot\alpha}}}\right|_{X}\right)\,,
\ee
using the incidence relations. Clearly, the first term in the parentheses vanishes, since $\lambda_{\alpha}\lambda^{\alpha}=0$. So if our twistor representative $f$ is \emph{holomorphic} (i.e., does not depend on the complex conjugated twistor variables), then it seems that our integral formula does indeed obey the z.r.m. equation. In terms of the complex structure on $\PT$, this holomorphicity condition can be phrased as: $\dbar f=0$.

\medskip

In the above argument, we have been a bit fast-and-loose, failing to specify what exactly we mean by the anti-holomorphic dependence on twistor space. As we learned in the previous lecture, to be precise about this, we must specify some reality conditions on twistor space. For concreteness, let's go through the calculation again, now with the explicit choice of Euclidean reality conditions on twistor space. 

Since $f$ is a $(0,1)$-form on $\PT$, we can expand it in the basis \eqref{T*bar}:
\be\label{nhE1}
f=f_{0}\,\bar{e}^{0}+f_{\dot\beta}\,\bar{e}^{\dot\beta}\,.
\ee
In the integral formula \eqref{nhPT3}, it is clear that only the first of these terms appears in the restriction $f|_{X}$, since $\bar{e}^{\dot\beta}$ does not point along the $\CP^1$-fibre direction of the Euclidean twistor space. So \eqref{nhPT3} can be written as
\be\label{nhE2}
\phi_{\alpha_1\cdots\alpha_{2s}}(x)=\int_{X}\la\lambda\,\d\lambda\ra\wedge \lambda_{\alpha_1}\cdots\lambda_{\alpha_{2s}}\,f_{0}|_{X}\,\bar{e}^0\,.
\ee
Using the basis \eqref{Tbar}, we can now compute the derivative:
\be\label{nhE3}
\partial^{\alpha_{1}\dot\alpha}\,\phi_{\alpha_1\cdots\alpha_{2s}}=\int_{X}\la\lambda\,\d\lambda\ra\wedge\lambda_{\alpha_2}\cdots\lambda_{\alpha_{2s}}\,\dbar^{\dot\alpha}f_{0}|_{X}\,\bar{e}^{0}=\int_{X}\omega\,\lambda_{\alpha_2}\cdots\lambda_{\alpha_{2s}}\,\dbar^{\dot\alpha}f_{0}|_{X}\,,
\ee
where
\be\label{Kahler}
\omega=\la\lambda\,\d\lambda\ra\wedge\bar{e}^{0}=\frac{\la\lambda\,\d\lambda\ra\wedge\la\hat{\lambda}\,\d\hat{\lambda}\ra}{\la\lambda\,\hat{\lambda}\ra^2}\,,
\ee
is the volume form on $\CP^1$.

At this point, we have only used the fact that $f$ is a $(0,1)$-form on $\PT$ of weight $-2s-2$. Now we can consider the action of the complex structure $\dbar$ on $f$:
\be\label{nhE4}
\dbar f=\left(\bar{e}^{0}\,\dbar_{0}+\bar{e}^{\dot\alpha}\,\dbar_{\dot\alpha}\right) \left(f_{0}\,\bar{e}^{0}+f_{\dot\beta}\,\bar{e}^{\dot\beta}\right)=\left(\dbar_{0}f_{\dot\alpha}-\dbar_{\dot\alpha}f_{0}\right)\bar{e}^{0}\wedge\bar{e}^{\dot\alpha} + \dbar_{\dot\alpha}f_{\dot\beta}\,\bar{e}^{\dot\alpha}\wedge\bar{e}^{\dot\beta}\,.
\ee
If $\dbar f=0$, then the terms proportional to $\bar{e}^{0}\wedge\bar{e}^{\dot\alpha}$ and $\bar{e}^{\dot\alpha}\wedge\bar{e}^{\dot\beta}$ must vanish independently, since these are distinct $(0,2)$-forms on twistor space. Thus, the condition $\dbar f=0$ imposes
\be\label{nhE5}
\dbar_{0}f_{\dot\alpha}=\dbar_{\dot\alpha}f_{0}\,,
\ee
on the components of $f$.

Feeding this back into \eqref{nhE3}, we find that
\be\label{nhE6}
\partial^{\alpha_{1}\dot\alpha}\,\phi_{\alpha_1\cdots\alpha_{2s}}=\int_{X}\omega\,\lambda_{\alpha_2}\cdots\lambda_{\alpha_{2s}}\,\dbar_{0}f^{\dot\alpha}|_{X} =\int_{X}\dbar_{0}\left(\omega\,\lambda_{\alpha_2}\cdots\lambda_{\alpha_{2s}}\,f^{\dot\alpha}|_{X}\right)=0\,,
\ee
which vanishes as a total derivative on the Riemann sphere $X\cong\CP^1$. (You might worry that the second equality in \eqref{nhE6} is missing some terms, but you can easily check that $\dbar_0 \omega=0$.) So sure enough, the condition $\dbar f=0$ imposes that $\phi_{\alpha_1\cdots\alpha_{2s}}$ obeys the z.r.m. equation. 

\medskip

The space of $f$s which obey $\dbar f=0$ contains some trivial solutions to the z.r.m. equations which we would like to get rid of, though. Since $\dbar^2=0$, it follows that any $f$ which can be written as $f=\dbar g$, for some $g\in\Omega^{0}(\PT,\cO(-2s-2))$ will automatically obey $\dbar f=0$. By an argument identical to the one used above, you can convince yourself that any such $f$ actually leads to a vanishing space-time field (i.e., $\phi_{\alpha_1\cdots\alpha_{2s}}=0$). Thus, it seems that the space of representatives on twistor space we want to consider is actually
\be\label{PenT1}
\left\{f\in\Omega^{0,1}(\PT,\cO(-2s-2)) \mbox{ which obey } \dbar f=0 \mbox{ and } f\neq\dbar g\right\}\,.
\ee
Such spaces of differential forms are well-studied objects in differential and algebraic geometry (which you may have encountered in other physics contexts), known as \emph{cohomology groups}. In particular, the set \eqref{PenT1} is the (Dolbeault) cohomology group denoted $H^{0,1}(\PT,\,\cO(-2s-2))$. You should read this notation as: the set of $(0,1)$-forms on $\PT$ of weight $-2s-2$ which obey $\dbar f=0$ and cannot be written as $f=\dbar g$. An element of a cohomology group is often referred to as a `cohomology class.'\footnote{For those who have been exposed to cohomology before, this is another place where we see that it was crucial for $\PT$ to be an open subset of $\CP^3$ rather than the entire projective space: these cohomology groups are empty for $\CP^3$! Physically, this is the statement that to have interesting solutions to the wave equation we need a non-compact space-time.}

So we have established that negative helicity z.r.m. fields on $\M_\C$ can be specified by twistor cohomology classes. It is straightforward to do something similar for z.r.m. fields of non-negative helicity as well (we will write the corresponding integral formulae momentarily). It turns out that this relationship also goes the other way: \emph{every} z.r.m. field on $\M_\C$ (which is suitably smooth) can be represented by a twistor cohomology class of a certain weight/homogeneity. Proving this other direction is a bit more technical, but if you are interested then you can look at the proof in~\cite{Eastwood:1981jy}.

The result is an isomorphism, known as the \emph{Penrose transform}:
\be\label{PenT2}
\left\{\mbox{helicity } h \mbox{ z.r.m. fields on } \M_{\C}\right\}\cong H^{0,1}(\PT,\,\cO(2h-2))\,,
\ee
for $h$ any integer or half-integer. Given a cohomology class on twistor space, the corresponding z.r.m. field on space-time can be constructed by means of an integral formula. The negative helicity case we have already seen; the other two cases are similar:
\be\label{PThn}
h<0 \qquad \phi_{\alpha_{1}\cdots\alpha_{2|h|}}(x)=\int_{X}\la\lambda\,\d\lambda\ra\wedge\lambda_{\alpha_1}\cdots\lambda_{\alpha_{2|h|}}\,f|_{X}\,,
\ee
\be\label{PThz}
h=0 \qquad \phi(x)=\int_{X}\la\lambda\,\d\lambda\ra\wedge f|_{X}\,,
\ee
\be\label{PThp}
h>0 \qquad \tilde{\phi}_{\dot\alpha_{1}\cdots\dot\alpha_{2h}}(x)=\int_{X}\la\lambda\,\d\lambda\ra\wedge\left.\frac{\partial}{\partial\mu^{\dot\alpha_1}}\cdots\frac{\partial}{\partial\mu^{\dot\alpha_{2h}}}\,f\right|_{X}\,.
\ee
You can readily check that the $h\geq 0$ integral formulae obey the z.r.m. equations by using holomorphicity and the incidence relations. Given a z.r.m. field on $\M_{\C}$ there is not, in general, a canonical way to reconstruct the twistor representative; this is partially due to the large redundancy of adding `gauge transformations' $\dbar g$ to any twistor representative, which does not change the cohomology class. In Euclidean signature, there is a canonical way to construct twistor representatives for some z.r.m. fields due to Woodhouse~\cite{Woodhouse:1985id}.\footnote{This is an interesting and useful procedure, which we do not have the time to cover here, but Woodhouse's paper is readable and you should be able to understand the necessary sections with the material covered in the lectures up to this point!}

\subsubsection*{\textit{The Sparling transform}}

We have already argued that z.r.m. fields are natural objects to study when talking about massless free fields: they are gauge invariant and manifest the positive/negative helicity decomposition in four-dimensions in terms of the SD/ASD decomposition of linearized curvature tensors. Suppose, however, that you really wanted to recover the gauge \emph{potential} associated with a z.r.m. field. Is there a way to do this directly from the twistor data? In the positive helicity case ($h>0$), there is a nice construction which allows us to do this due to Sparling~\cite{Sparling:1990}.

Let's consider the $h=+1$ case; we want to find a way to construct a space-time Maxwell field $A_{a}(x)$ from a Penrose transform representative on twistor space. By \eqref{PenT2}, we know that the twistor representative for a positive helicity Maxwell field is a cohomology class
\be\label{SparT1}
a\in H^{0,1}(\PT,\,\cO)\,.
\ee
Consider the restriction of this representative to a line $X\subset\PT$ corresponding to a point in $\M_{\C}$. Since $a$ is a cohomology class on $\PT$, it is also a cohomology class on $X$:
\be\label{SparT2}
a|_{X}\in H^{0,1}(X,\,\cO)\cong H^{0,1}(\CP^1, \cO)\,.
\ee
However, the cohomology group $H^{0,1}(\CP^1, \cO)$ is actually empty.

There's a fairly intuitive way to see why this is the case. First, let's try to construct an element of $\Omega^{0,1}(\CP^1,\cO)$: this will be a $(0,1)$-form on the Riemann sphere which is homogeneous of weight zero. With the standard complex structure on $\CP^1$, such an object must be proportional to $\la\hat{\lambda}\,\d\hat{\lambda}\ra$, which has weight $+2$ in $\hat{\lambda}_{\alpha}$. So to form a homogeneous $(0,1)$-form, we need an object which looks like
\begin{equation*}
 \frac{\la\hat{\lambda}\,\d\hat{\lambda}\ra\,\la a\,b\ra}{\la a\,\hat{\lambda}\ra\, \la b\,\hat{\lambda}\ra}\,,
\end{equation*}
for $a_{\alpha},b_{\alpha}$ the homogeneous coordinates of some fixed points on $\CP^1$. But such an object is clearly not holomorphic on $\CP^1$, and so cannot be a cohomology class. (If you know some algebraic geometry, you can easily prove that $H^{0,1}(\CP^1,\cO)=\emptyset$ using Serre duality or the Riemann-Roch theorem.)

So if $a|_{X}\in H^{0,1}(\CP^1,\,\cO)$ and $H^{0,1}(\CP^1,\cO)=\emptyset$, it follows that $a|_{X}$ must trivially obey $\dbar|_{X} a|_{X}=0$:
\be\label{SparT3}
a|_{X}=\dbar|_{X}\,h(x,\lambda,\hat{\lambda})\,,
\ee
for some function $h$ which is homogeneous of degree zero in $\lambda,\hat{\lambda}$. Furthermore, since $a$ is defined on $\PT$, it can depend on $x^{\alpha\dot\alpha}$ only through the combination $x^{\alpha\dot\alpha}\lambda_{\alpha}$ (or its complex conjugate). This is just the usual statement of the incidence relations, and implies
\be\label{SparT4}
\dbar|_{X}\left(\lambda^{\alpha}\partial_{\alpha\dot\alpha} h\right)=\lambda^{\alpha}\partial_{\alpha\dot\alpha} a|_{X}=0\,.
\ee
This means that $\lambda^{\alpha}\partial_{\alpha\dot\alpha} h$ is a function of $x$ and $(\lambda,\hat{\lambda})$ which is holomorphic and of weight $+1$ in $\lambda$. It is clear (by an extension of Liouville's theorem), that any such function must take the form:
\be\label{SparT5}
\lambda^{\alpha}\partial_{\alpha\dot\alpha} h(x,\lambda,\hat{\lambda}) = \lambda^{\alpha}\,A_{\alpha\dot\alpha}(x)\,.
\ee
This $A_{\alpha\dot\alpha}(x)$ is precisely the Maxwell potential we set out to find. A similar story works for other positive helicity fields of higher spin (e.g., the linearized gravity case is worked out in~\cite{Mason:2008jy}); you may find it instructive to try this for yourself.

\subsubsection*{Exercise: \textit{momentum eigenstates}}

When we do Feynman diagram calculations in perturbative QFT, we usually take the wavefunctions of our external states to be modeled on exponential plane waves, $\e^{\im k\cdot x}$, for $k_{a}$ an on-shell momentum. In the massless case ($k^2=0$), we know that we can represent this $k_{a}\leftrightarrow p_{\alpha}\,\tilde{p}_{\dot\alpha}$. This exercise is concerned with how to construct twistor representatives for such states via the Penrose transform.

\begin{enumerate}
 \item \emph{Holomorphic delta functions}: Let $z$ be the usual complex coordinate on $\C$, and consider
 \be\label{holdelta}
  \bar{\delta}(z):=\frac{1}{2\pi \im}\,\d\bar{z}\,\frac{\partial}{\partial\bar{z}}\left(\frac{1}{z}\right)=\frac{1}{2\pi\im}\,\dbar\left(\frac{1}{z}\right)\,.
 \ee
 Show that this object acts like a holomorphic analogue of the Dirac delta function by integrating against a holomorphic test function, $f(z)$. In particular, show that
 \begin{equation*}
  \int_{D} \d z\wedge\bar{\delta}(z)\,f(z)=f(0)\,,
 \end{equation*}
 where $D\subset\C$ is a disc with boundary $\Gamma$ enclosing the origin.
 
 Let $\bar{\delta}^{2}(\lambda_{\alpha})$ be the natural extension of the holomorphic delta function to 2-spinor quantities:
 \begin{equation*}
  \bar{\delta}^{2}(\lambda_{\alpha}):=\bigwedge_{\alpha=0,1} \dbar\left(\frac{1}{\lambda_{\alpha}}\right)\,.
 \end{equation*}
 Clearly, $\bar{\delta}^{2}(\lambda_{\alpha})$ should be interpreted as a sort of $(0,2)$-form which has support only where its argument vanishes. Show that
 \begin{equation*}
  \int_{\C^*}\frac{\d s}{s^{2h-1}}\,\bar{\delta}^{2}(p_{\alpha}-s\,\lambda_{\alpha})=\left(\frac{\la a\,\lambda\ra}{\la a\,p\ra}\right)^{2h-1}\,\bar{\delta}(\la\lambda\,p\ra)\,,
 \end{equation*}
 where $p_{\alpha},a_{\alpha}$ are constant 2-spinors which obey $\la p\,a\ra\neq0$. It looks like the RHS of this equations depends on a spinor ($a_{\alpha}$) which doesn't appear on the LHS...why is this not a problem?
 
 \item \emph{Twistor representatives}: Consider
 \be\label{momeig1}
 f^{[h]}(Z)=\int_{\C^*}\frac{\d s}{s^{2h-1}}\,\bar{\delta}^{2}(p_{\alpha}-s\,\lambda_{\alpha})\,\exp\left(\im\,s\,[\mu\tilde{p}]\right)\,,
 \ee
 where $p_{\alpha},\tilde{p}_{\dot\alpha}$ are constant 2-spinors. Show that $f^{[h]}\in H^{0,1}(\PT,\cO(2h-2))$. (Hint: treat the parameter $s$ as a scaling parameter with weight $-1$ with respect to the projective scale on twistor space, or use the result you proved above.)
 
 \item \emph{Penrose transform}: Using the integral formulae \eqref{PThn} -- \eqref{PThp}, show that $f^{[h]}$ gives rise to the following momentum eigenstate z.r.m. fields on space-time:
 \begin{equation*}
  (h<0) \quad p_{\alpha_1}\cdots p_{\alpha_{2|h|}}\,\e^{\im\,k\cdot x}\,, \qquad (h=0) \quad \e^{\im\,k\cdot x}\,, \qquad (h>0) \quad \tilde{p}_{\dot\alpha_1}\cdots\tilde{p}_{\dot\alpha_{2h}}\,\e^{\im\,k\cdot x}\,,
 \end{equation*}
 where $k_{\alpha\dot\alpha}=p_{\alpha}\tilde{p}_{\dot\alpha}$.
 
 \item \emph{Sparling transform}: Let $h=+1$ in \eqref{momeig1}. Perform the Sparling transform on $f^{[1]}$ to obtain a space-time gauge field. You will need to manipulate expressions along the way, using the Schouten identity ($\la a\,b\ra\la c\,d\ra = \la a\,c\ra\la b\,d\ra + \la a\,d\ra\la c\,b\ra$) and dropping any terms which vanish on the support of the holomorphic delta functions. You should find
 \begin{equation*}
  h(x,\lambda,\hat{\lambda})=\frac{\la a\,\lambda\ra}{\la a\,p\ra\,\la\lambda\,p\ra}\,\e^{\im\,k\cdot x}\,, \qquad A_{\alpha\dot\alpha}(x)=\frac{a_{\alpha}\,\tilde{p}_{\dot\alpha}}{\la a\,p\ra}\,\e^{\im\,k\cdot x}\,.
 \end{equation*}
 Show that different choices of the spinor $a_{\alpha}$ correspond to gauge transformations of $A_{\alpha\dot\alpha}(x)$.

\end{enumerate}

%% file: TLecture4.tex
\section{Gauge Theory in Twistor Space}

The Penrose transform gives us a way to study massless free fields in Minkowski space in terms of twistor data. Of course, to study interesting physical problems with twistor theory we need to be able to describe non-linear, or interacting, field theories. In this lecture, we consider a familiar interacting field theory with obvious applicability to physics: non-abelian Yang-Mills theory. We will see that twistor theory provides a natural description of a non-linear, integrable subsector of Yang-Mills theory, which can be used to build up a twistor description of the full theory which is perturbatively equivalent to the space-time formulation.


\subsection{The Ward correspondence}

What is the natural language to talk about gauge theory on twistor space? To answer this question, it is instructive to first think about the natural language for gauge fields in space-time. This is done by introducing a \emph{gauge field}, which we usually talk about in terms of a 1-form $A_{a}(x)$, which takes values in the adjoint representation of the gauge group, $G$. We denote the (complexified) Lie algebra of the gauge group by $\mathfrak{g}$. The physics of the gauge field arises by modifying the natural derivative structure on space-time -- namely, the coordinate derivative $\partial_{a}$ -- to include the gauge field $\partial_{a}\rightarrow D_{a}=\partial_{a}+A_{a}$. The resulting derivative operator, $D_{a}$, is often referred to as the \emph{gauge connection}.

The natural objects on which the gauge connection acts are functions or tensors which are also valued in representations of the gauge group. In particular, if $f(x)$ is valued in the fundamental and $\Phi(x)$ is valued in the adjoint representation of $G$, then the gauge connection acts as
\begin{equation*}
 D_{a} f= \partial_{a} f + A_{a}\, f\,, \qquad D_a \Phi=\partial_{a}\Phi + [A_{a},\,\Phi]\,,
\end{equation*}
where $[\cdot,\,\cdot]$ is Lie bracket, which is simply the commutator between $\mathfrak{g}$ matrices. We know that the gauge field $A_{a}$ itself is not invariant; the physical information encoded in a gauge theory should be invariant under gauge transformations. These are just shifts of the gauge field by adjoint-valued functions:
\be\label{stgt1}
A_{a}\rightarrow \mathsf{g}(x)\,A_{a}\, \mathsf{g}^{-1}(x) - \partial_{a}\mathsf{g}(x)\, \mathsf{g}^{-1}(x)\,,
\ee
with $\mathsf{g}(x)$ valued in $\mathfrak{g}$. It is straightforward to see that the field strength,
\be\label{stgt2}
F_{ab}=[D_{a},\,D_{b}]=\partial_{a} A_{b} - \partial_{b}A_{a} + [A_{a},\,A_{b}]\,,
\ee
transforms covariantly under these gauge transformations: $F_{ab}\rightarrow \mathsf{g} F_{ab} \mathsf{g}^{-1}$. So (local) gauge-invariant quantities can be formed by taking traces of combinations of the field strength.

\medskip

By analogy, we should look to formulate gauge theory in twistor space by deforming the natural differential structure on $\PT$ by an adjoint-valued connection. As we have learned, the natural differential structure on twistor space is the complex structure, in the form of the operator $\dbar$. Therefore, the twistor space version of a gauge connection is a deformed complex structure, which looks locally like:
\be\label{Tgc1}
\bar{D}=\dbar + a\,, \qquad a\in\Omega^{0,1}(\PT,\,\mathfrak{g})\,.
\ee
In other words, the natural gauge field on twistor space is a $(0,1)$-form taking values in the adjoint of the gauge group. The operator $\bar{D}$ is called a covariant almost complex structure, a $(0,1)$-connection, or a partial connection. These names can be used interchangeably; they all reflect the fact that the natural notion of a gauge connection on $\PT$ is a deformation of the standard complex structure.

Just as gauge covariant information is packaged in the field strength $F_{ab}$ on space-time, gauge-covariant information is packaged in the curvature of $\bar{D}$ on twistor space. This is a $(0,2)$-form, referred to as the `anti-holomorphic curvature' of the partial connection:
\be\label{Tgc2}
F^{(0,2)}=[\bar{D},\,\bar{D}]\in\Omega^{0,2}(\PT,\,\mathfrak{g})\,.
\ee
Under a gauge transformation
\be\label{Twistorgt}
\bar{D}\rightarrow \gamma(Z)\,\bar{D} \gamma^{-1}(Z)\,, \qquad \gamma\in\Omega^{0}(\PT,\,\mathfrak{g})\,,
\ee
you can show that $F^{(0,2)}\rightarrow\gamma F^{(0,2)}\gamma^{-1}$, as expected. 

A proper geometric treatment of partial connections entails the use of fibre bundles. If you are already familiar with these concepts, then it's probably clear to you what the general setup should be. If not, then don't worry: even the simplest possible example captures all of essential features from the perspective of twistor theory. We say that $E\rightarrow \PT$ is a rank $N$ vector bundle over twistor space if it looks locally like $E\cong\C^{N}\times\PT$; its fibre over a point $Z\in\PT$ is just a copy of the $N$-dimensional vector space: $E|_{Z}\cong\C^N$. We will demand that when $E$ is restricted to a line $X\cong\CP^1$ in twistor space, it is trivial: $E|_{X}\cong\C^{N}\times X$ (or in the language of Chern classes, $c_{1}(E|_X)=0$). This latter requirement will means that information encoded in this vector bundle can be translated to local information on space-time.

Its easy to see that $\bar{D}$ is best thought of as a connection on the vector bundle $E$ itself. The endomorphisms of the fibres of $E$ are just $N\times N$ complex matrices, so it follows that $\End(E)\cong\mathfrak{gl}(N,\C)$. Thus, the rank $N$ vector bundle $E$ naturally encodes the gauge transformations associated with gauge group $G=\GL(N,\C)$. As we will see later, other gauge groups arise by endowing $E$ with additional structures.

\medskip

Having established that the natural analogue of a gauge field on $\PT$ is the partial connection $\bar{D}$ on a rank $N$ vector bundle, we can ask what sort of field equations can be imposed on the partial connection. Any reasonable field equation should be gauge invariant, which means that it must be phrased in terms of the anti-holomorphic curvature $F^{(0,2)}$. We can't impose the usual Yang-Mills equations, because the partial connection only points in the anti-holomorphic directions of twistor space. Instead, we can simply consider the field equation $F^{(0,2)}=0$; this is the condition for the vector bundle $E$, equipped with partial connection $\bar{D}$, to be \emph{holomorphic}. Equivalently, this means that $\bar{D}^2=0$ and thus defines an integrable covariant complex structure.

To see precisely what the equation $F^{(0,2)}=0$ entails, it's helpful to pick a reality structure to do our calculations in. As usual, we'll take the Euclidean reality structure, where we can use the bases \eqref{Tbar} and \eqref{T*bar}. This means that we can expand the twistor gauge field as
\be\label{WardT1}
a=a_{0}\,\bar{e}^{0} + a_{\dot\alpha}\,\bar{e}^{\dot\alpha}\,,
\ee
where the coefficients $\{a_0,\,a_{\dot\alpha}\}$ are adjoint-valued functions on $\PT$, homogeneous of weight $+2$ and $+1$ respectively. We can then compute
\be\label{WardT2}
F^{(0,2)}=\left(\dbar_{0}a_{\dot\alpha}-\dbar_{\dot\alpha} a_{0}-\left[a_{\dot\alpha},\,a_{0}\right]\right)\bar{e}^{0}\wedge\bar{e}^{\dot\alpha} + \left(\dbar_{\dot\alpha}a_{\dot\beta} +\left[a_{\dot\alpha},\,a_{\dot\beta}\right]\right)\bar{e}^{\dot\alpha}\wedge\bar{e}^{\dot\beta}\,.
\ee
Note that all contributions to $F^{(0,2)}$ from the component $a_0$ are given by
\be\label{a0gf}
\left(\dbar_{\dot\alpha} a_{0}+\left[a_{\dot\alpha},\,a_{0}\right]\right)\bar{e}^{\dot\alpha}\wedge\bar{e}^{0}=\bar{D} (a_{0}\,\bar{e}^0)\,,
\ee
which means that $a_{0}$ can be removed by a gauge transformation. 

There is another nice way of seeing this. The gauge freedom \eqref{Twistorgt} can be used to impose $\dbar|_{X}^{*} a_{0}=0$ on each $X\cong\CP^1$ in twistor space, where $\dbar_{X}^{*}$ is the adjoint operator of $\dbar|_{X}=\bar{e}^{0}\dbar_0$. Now, $a_{0}$ is the component of a $(0,1)$-form on $\CP^1$, and as such it must obey $\dbar|_{X}a_{0}=0$ (since there are no $(0,2)$-forms on $\CP^1$). So this choice of gauge actually forces $a_0$ to be a harmonic function on $X$: $\dbar|_{X}^{*}\dbar|_{X}a_{0}=0$. The Hodge theorem tells us that every harmonic function corresponds to a cohomology class, so
\be\label{a0gf1}
a|_{X}=a_{0}\,\bar{e}^{0}\in H^{1}(\CP^1,\,\mathfrak{gl}(N,\C))\,.
\ee
As we already saw in the previous lecture, this cohomology group is actually empty: $H^{1}(\CP^1,\,\mathfrak{gl}(N,\C))=\emptyset$. Thus, we can consistently set $a_{0}=0$ as a gauge condition.

With this choice, the gauge field on twistor space becomes $a=a_{\dot\alpha}\bar{e}^{\dot\alpha}$, and the anti-holomorphic curvature is given by
\be\label{WardT3}
F^{(0,2)}=\dbar_{0}a_{\dot\alpha}\,\bar{e}^{0}\wedge\bar{e}^{\dot\alpha} + \left(\dbar_{\dot\alpha}a_{\dot\beta} +\left[a_{\dot\alpha},\,a_{\dot\beta}\right]\right)\bar{e}^{\dot\alpha}\wedge\bar{e}^{\dot\beta}\,.
\ee
Imposing the field equation $F^{(0,2)}=0$ is therefore equivalent to two equations on the remaining components of $a$:
\be\label{WardT4}
\dbar_{0}a_{\dot\alpha}=0\,, \qquad \dbar_{[\dot\alpha} a_{\dot{\beta}]}+[a_{\dot\alpha},\,a_{\dot\beta}]=0\,.
\ee
The first of these equations tells us that $a_{\dot\alpha}(x,\lambda,\hat{\lambda})$ is holomorphic as a function of $(\lambda,\hat{\lambda})$. We encountered this situation in the previous lecture in the context of the Sparling transform; by Liouville's theorem, it follows that
\be\label{WardT5}
\dbar_{0}a_{\dot\alpha}=0 \quad \Rightarrow \quad a_{\dot\alpha}(x,\lambda,\hat{\lambda})=\lambda^{\alpha}\,A_{\alpha\dot\alpha}(x)\,,
\ee
where $A_{\alpha\dot\alpha}(x)$ is valued in $\mathfrak{gl}(N,\C)$. So the first equation in \eqref{WardT4} tells us that the holomorphic partial connection on $\PT$ encodes a gauge field on $\R^4$. 

Clearly, the second equation of \eqref{WardT4} will impose some further conditions on this space-time gauge field. Plugging \eqref{WardT5} into this second equation, we find that
\be\label{WardT6}
\dbar_{[\dot\alpha} a_{\dot{\beta}]}+[a_{\dot\alpha},\,a_{\dot\beta}]=\epsilon_{\dot\alpha\dot\beta}\,\lambda^{\alpha}\lambda^{\beta}\left(\partial_{\alpha\dot\gamma}A_{\beta}^{\dot\gamma}+\left[A_{\alpha\dot\gamma},\,A_{\beta}^{\dot\gamma}\right]\right)=\epsilon_{\dot\alpha\dot\beta}\,\lambda^{\alpha}\lambda^{\beta}\,F_{\alpha\beta}=0\,,
\ee
where $F_{\alpha\beta}$ is the anti-self-dual portion of the field strength of the gauge field. This equation can only be satisfied for non-trivial connections if $F_{\alpha\beta}=0$ -- that is, if the gauge field on $\R^4$ is \emph{self-dual}.

In summary, we have shown that every holomorphic rank $N$ vector bundle on $\PT$ (i.e., a partial connection $\bar{D}$ on $E\rightarrow\PT$ obeying $F^{(0,2)}=0$) leads to a self-dual Yang-Mills field on $\R^4$ with gauge group $\GL(N,\C)$. These SD gauge fields on $\R^4$ are known as Yang-Mills \emph{instantons}.

\medskip

One can naturally ask if this correspondence works the other way around. That is, suppose we are given a $\GL(N,\C)$ gauge field on space-time which is self-dual: $F_{\alpha\beta}=0$. Does this define a holomorphic, rank $N$ vector bundle on twistor space? It is easy to see that this is so; indeed, we can construct the corresponding holomorphic bundle over every point of $\PT$ for complexified space-time and impose reality conditions at the end of this construction.

Our starting point is a SD gauge field on $\M_\C$; this has a field strength:
\be\label{cWardT1}
F_{ab}=\epsilon_{\alpha\beta}\,\tilde{F}_{\dot\alpha\dot\beta}\,,
\ee
by virtue of the SD condition. Every point $Z\in\PT$ corresponds to an $\alpha$-plane in $\M_\C$; recall that this is a totally null 2-plane in $\M_\C$ whose tangent vectors are all proportional to $\lambda^{\alpha}$. Consider the restriction of the field strength to any such $\alpha$-plane; this is given by
\be\label{cWardT2}
F_{ab}|_{\alpha-\mathrm{plane}}=v^{a}\,w^{b}\,F_{ab}\,,
\ee
where $v^a$, $w^b$ are any two tangent vectors to the $\alpha$-plane. By definition, $v^{a}=\lambda^{\alpha}\tilde{v}^{\dot\alpha}$, $w^{b}=\lambda^{\beta}\tilde{w}^{\dot\beta}$ for some spinors $\tilde{v}^{\dot\alpha},\tilde{w}^{\dot\beta}$, so we find
\be\label{cWardT3}
F_{ab}|_{\alpha-\mathrm{plane}}=\tilde{v}^{\dot\alpha}\tilde{w}^{\dot\beta}\,\lambda^{\alpha}\lambda^{\beta}\,\epsilon_{\alpha\beta}\,\tilde{F}_{\dot\alpha\dot\beta}=0\,.
\ee
In other words, SD gauge fields are flat upon restriction to $\alpha$-planes.

This means that the space of covariantly constant functions valued in the fundamental representation on the $\alpha$-plane is equivalent to the space of constant functions. So to each $\alpha$-plane we can assign a vector space
\be\label{cWardT4}
E|_{Z}=\left\{\left.\mathfrak{s}(x) \mbox{ valued in } \C^N \right|\, D_{a}\mathfrak{s}|_{\alpha-\mathrm{plane}}=0\right\}\cong\C^{N}\,.
\ee
In particular, we can associate a copy of $\C^N$ to every point $Z\in\PT$ in this way. It is easy to see that this leads to a rank $N$ vector bundle over $\PT$ which is topologically trivial upon restriction to lines in twistor space. Furthermore, since this is a totally holomorphic construction, the resulting vector bundle is holomorphic.

\medskip

This establishes a one-to-one correspondence between Yang-Mills instantons with gauge group $\GL(N,\C)$ on $\M_\C$ and rank $N$ holomorphic vector bundles $E\rightarrow\PT$ satisfying $E|_{X}\cong\C^{N}\times\CP^1$. Known as the \emph{Ward correspondence}~\cite{Ward:1977ta}, it constitutes one of the most important results from the early years of twistor theory. The Ward correspondence is easily extended to any gauge group by imposing further conditions on the holomorphic vector bundle on twistor space. For example, SU$(N)$ instantons are described by requiring $E\rightarrow\PT$ to be equipped with a positive real form, and the determinant line bundle $\det(E)$ to be trivial. These structures enable the construction of a Killing form and ensure that the transition matrices of $E$ are unimodular, respectively. 

The Ward correspondence has been extremely influential in the study of classical integrable systems. It led to early constructions of Yang-Mills instantons~\cite{Atiyah:1977pw} and was a major influence on the ADHM construction of all Yang-Mills instantons~\cite{Atiyah:1978ri}. Furthermore, myriad integrable systems in lower dimensions such as the Bogomolny monopole equations in $d=3$~\cite{Ward:1981jb,Hitchin:1982gh}, Hitchin systems in $d=2$~\cite{Hitchin:1986vp,Hitchin:1987mz}, and even the non-linear Schr\"odinger and Kortweg-de Vries equations~\cite{Mason:1989kk,Mason:1992vd} can be viewed as symmetry reductions of the instanton equations which have twistor constructions via the Ward correspondence.

There is also a gravitational analogue of the Ward correspondence, known as the \emph{non-linear graviton} construction~\cite{Penrose:1976js,Newman:1976gc,Atiyah:1978wi}. This gives a one-to-one correspondence between complex deformations of twistor space and four-dimensional complex space-times with a self-dual conformal (holomorphic) metric. By this, we mean that the complex structure of the deformed twistor space defines, up to conformal equivalence, a space-time metric whose Weyl tensor obeys $\Psi_{\alpha\beta\gamma\delta}=0$. The conformal class can also be fixed to a SD Einstein metric by including some extra data on the twistor space (namely, a `weighted contact structure')~\cite{Ward:1980am}. Although we won't have time to discuss the non-linear graviton construction in these lectures, you can intuitively imagine it as the Ward correspondence with the holomorphic vector bundle $E$ on $\PT$ replaced by the tangent bundle $T_{\PT}$ itself.


\subsection{Perturbative expansion around the self-dual sector}

Although the instanton sector is important, it is a long way from the full interacting Yang-Mills theory. Indeed as a QFT, self-dual Yang-Mills theory isn't very interesting: it is classically integrable (indeed, the Ward correspondence demonstrates this), non-unitary and `almost' free. This last fact can be seen by looking at the perturbative scattering amplitudes of the theory: the only non-vanishing amplitudes are at tree-level (for one negative helicity and two positive helicity external gluons) and at one-loop (for all positive helicity external gluons). Can we get a twistor description of full Yang-Mills theory?

Trying to find an answer to this question was one of the major problems for twistor theory during the 1980s, and became known as the `googly problem,' a moniker derived from a certain kind of ball which can be bowled in cricket. The essence of the googly problem for Yang-Mills theory is trying to find a twistor description of general Yang-Mills field configurations. To date, there is still no (fully non-linear) solution to the googly problem, despite decades of work by a hard-core of twistor theorists on the subject.\footnote{On a rainy day, you can amuse yourself by looking through the archives of \textit{Twistor Newsletter} (an in-house journal published by twistor theorists at Oxford from 1976-2000) to get a feel for the sort of solutions which have been attempted in the past: \texttt{http://people.maths.ox.ac.uk/lmason/Tn/} . More recently Penrose proposed another potential solution~\cite{Penrose:2015lla}, called `palatial twistor theory,' but I think it's still unclear whether this actually solves the googly problem (and if so, in a useful way).} 

You might worry that this is the end of the story, but it turns out that a \emph{perturbative} solution to the googly problem can be found which is good enough for computing many quantities of interest from the perspective of perturbative QFT. As we will see, this provides an alternative description of gauge theory in terms of a perturbative expansion around the SD sector, which is naturally amenable to twistor theory.

\medskip

The standard Yang-Mills action in flat space is given by
\be\label{YMa1}
S[A]=-\frac{1}{2\, g^2}\int \tr\!\left(F\wedge * F\right) = -\frac{1}{4\,g^2}\int \d^{4}x\,\tr\!\left(F_{ab}\,F^{ab}\right)\,,
\ee
where $g$ is the dimensionless coupling constant. Expanding the field strength into its self-dual and anti-self-dual parts, we find that
\be\label{YMa2}
S[A]=-\frac{1}{2\,g^2}\int \d^{4}x\,\tr\!\left(F_{\alpha\beta}\,F^{\alpha\beta}+\tilde{F}_{\dot\alpha\dot\beta}\,\tilde{F}^{\dot\alpha\dot\beta}\right)\,.
\ee
So far we haven't done anything fancy: \eqref{YMa2} is just the Yang-Mills action written in terms of the spinor decomposition of the field strength. 

Now, recall that the Yang-Mills action can be modified by the addition of the $\theta$-term:
\be\label{Theta}
\int \tr\!\left(F\wedge F\right) = 4 \int \d^{4}x\,\tr\!\left(\tilde{F}_{\dot\alpha\dot\beta}\,\tilde{F}^{\dot\alpha\dot\beta}-F_{\alpha\beta}\,F^{\alpha\beta}\right)\,.
\ee
While the presence of the $\theta$-term affects non-perturbative features of the gauge theory, it does not alter the perturbative physics in flat space-time since it is a topological term. Thus, we are free to add or subtract any multiple of \eqref{Theta} to the Yang-Mills action, and the result will still be perturbatively equivalent to Yang-Mills theory. In particular, let us add $\frac{1}{8g^2}$ times the $\theta$-term to the Yang-Mills action; this results in:
\be\label{YMa3}
S[A]+\frac{1}{8\,g^2}\int \tr\!\left(F\wedge F\right)=-\frac{1}{g^2}\int \d^{4}x\,\tr\!\left(F_{\alpha\beta}\,F^{\alpha\beta}\right)\,.
\ee
So this simplified action, which depends only on the ASD field strength of the gauge field, is perturbatively equivalent to Yang-Mills theory.

What have we gained by doing this? The answer is easier to see by introducing a Lagrange multiplier to re-express \eqref{YMa3}. Let $G_{\alpha\beta}(x)$ be symmetric in its spinor indices and valued in the adjoint of the gauge group, and consider the action:
\be\label{ChS1}
S[A,G]=\int \d^{4}x\,\tr\!\left(F_{\alpha\beta}\,G^{\alpha\beta}\right) +\frac{g^2}{4}\int \d^{4}x\,\tr\!\left(G_{\alpha\beta}\,G^{\alpha\beta}\right)\,.
\ee
The field equations of this action are:
\be\label{ChSfes}
F_{\alpha\beta}=-\frac{g^2}{2}\,G_{\alpha\beta}\,, \qquad D^{\alpha\dot\alpha}G_{\alpha\beta}=0\,,
\ee
from which it is easy to see that integrating out $G_{\alpha\beta}$ returns the action \eqref{YMa3}. The equations \eqref{ChSfes} are telling us something interesting in their own right, though. The ASD portion of the gauge field is encoded by $G_{\alpha\beta}$, which itself acts as a covariant z.r.m. field on-shell. When the coupling constant $g$ is vanishing, we recover the SD field equations: $F_{\alpha\beta}=0$. 

This means that advantage of working with the action \eqref{ChS1} -- which is perturbatively equivalent to the Yang-Mills action -- is that the coupling constant acts as a small parameter for perturbatively expanding around the SD sector of the theory. In other words, we have shown that Yang-Mills theory in Minkowski space admits a perturbative expansion around the SD (or instanton) sector -- something which is not at all obvious from the usual Yang-Mills action \eqref{YMa1}! 

This new formulation, often referred to as the \emph{Chalmers-Siegel} action, presents perturbative Yang-Mills theory in terms of ASD fluctuations around a non-linear SD background~\cite{Chalmers:1996rq}. From the perspective of twistor theory, this is just what we were hoping for: a perturbative solution to the googly problem. The Ward correspondence describes the non-linear SD sector, and the Penrose transform can be used to describe the ASD perturbations. As we will see, this means that the action \eqref{ChS1} can be lifted to twistor space.


\subsection{The twistor action} 

First, let's consider how to encode the purely SD sector of the action \eqref{ChS1} in twistor space. In terms of our new perturbative expansion, this is the zero-coupling limit, described on space-time by the action
\be\label{SDTA1}
S^{\mathrm{SD}}[A,G]=\int \d^{4}x\,\tr\!\left(F_{\alpha\beta}\,G^{\alpha\beta}\right)\,,
\ee
with field equations
\be\label{SDfes}
F_{\alpha\beta}=0\,,  \qquad D^{\alpha\dot\alpha}G_{\alpha\beta}=0\,.
\ee
By the Ward Correspondence, we know that the field equation $F_{\alpha\beta}=0$ is described on twistor space by a partial connection, $\bar{D}=\dbar+a$, which is holomorphic:
\be\label{SDTA2}
F_{\alpha\beta}=0 \quad \Leftrightarrow \quad F^{(0,2)}=[\bar{D},\,\bar{D}]=\dbar a+a\wedge a=0\,,
\ee 
with $a\in\Omega^{0,1}(\PT,\mathfrak{g})$ the twistor gauge connection. 

The field equation $F^{(0,2)}=0$ can be enforced dynamically on $\PT$ by using a Lagrange multiplier. Consider the action:
\be\label{SDTA}
S^{\mathrm{SD}}[a,g]=\int_{\PT}\D^{3}Z\wedge\tr\!\left[g\wedge\left(\dbar a+a\wedge a\right)\right]\,,
\ee
where $\D^{3}Z$ is the canonical holomorphic measure on $\CP^3$ of projective weight $+4$ given by
\be\label{tvol}
\D^{3}Z:=\epsilon_{ABCD}\,Z^{A}\,\d Z^{B}\wedge\d Z^{C}\wedge\d Z^{D}\,.
\ee
In order for this action to make sense as an integral over $\PT$, the Lagrange multiplier must be an adjoint-valued $(0,1)$-form on $\PT$, homogeneous of weight $-4$:
\be\label{tLM}
g\in\Omega^{0,1}(\PT,\,\cO(-4)\otimes\mathfrak{g})\,.
\ee
The field equations of the twistor action \eqref{SDTA} are thus
\be\label{T1fes}
\dbar a+a\wedge a=0\,, \qquad \bar{D} g=0\,,
\ee
the first of which is precisely the SD equation.

What about the second equation, $\bar{D}g=0$? On the support of the other field equation, $\bar{D}^2=0$, so the partial connection defines an integrable (covariant) complex structure on $\PT$. This means that on-shell, $g$ is in fact a cohomology class:
\be\label{SDTA3}
\bar{D}g=0 \Rightarrow g\in H^{0,1}_{\bar{D}}(\PT,\,\cO(-4)\otimes\mathfrak{g})\,.
\ee
Now, if we replaced $\bar{D}$ with the flat complex structure $\dbar$ and took the abelian gauge group $G=\U(1)$, then we could apply the Penrose transform to $g$, resulting in a z.r.m. field on space-time:
\be\label{abPT1}
G_{\alpha\beta}(x)=\int_{X}\la\lambda\,\d\lambda\ra\wedge \lambda_{\alpha}\lambda_{\beta}\,g|_{X}\,, \qquad \partial^{\alpha\dot\alpha} G_{\alpha\beta}=0\,.
\ee
So it seems that we get the correct twistor space field equation if a \emph{covariant, non-abelian} version of the Penrose transform holds.

As it turns out, this is the case. We'll leave part of the construction as an exercise at the end of the lecture, but even generalizing the integral formula for $G_{\alpha\beta}$ in terms of $g$ to the case of a non-abelian gauge group is a bit non-trivial. In particular, the partial connection $\bar{D}$ acts on a rank $N$ vector bundle $E\rightarrow \PT$; by assumption $E|_{X}$ is topologically trivial. However, it need not be \emph{holomorphically} trivial upon restriction to $X\cong\CP^1$. This means that we cannot \emph{a priori} compare fibres of the bundle holomorphically over two different points on a line in twistor space.

Now, $E|_{X}$ can be holomorphically trivialized if we can find a gauge transformation $\gamma(x,\lambda)$ for which 
\be\label{nabPT1}
\gamma(x,\lambda)\,\bar{D}|_{X}\,\gamma^{-1}(x,\lambda)=\dbar|_{X}\,,
\ee
that is, a gauge transformation which \emph{trivializes} the partial connection over each $X$. Intuitively, it's not hard to convince yourself that such a trivialization will exist perturbatively. Indeed, we imagine that we will always be using the action \eqref{SDTA} perturbatively -- that is, around `small' configurations of the twistor fields $a$ and $g$. If $a$ is `small', then $\bar{D}$ looks like $\dbar$, for which the partial connection is automatically holomorphically trivial.

Let $\gamma$ be this perturbatively constructed trivialization. Then the non-abelian version of the Penrose transform integral formula is given by:
\be\label{nabPT2}
G_{\alpha\beta}(x)=\int_{X}\la\lambda\,\d\lambda\ra\wedge\lambda_{\alpha}\lambda_{\beta}\,\gamma^{-1}(x,\lambda)\,g|_{X}\,\gamma(x,\lambda)\,.
\ee
With such an integral formula, you can show that the resulting $G_{\alpha\beta}$ is a covariant z.r.m. field on space-time provided $g$ is holomorphic with respect to the partial connection on twistor space.

\medskip

This establishes that $S^{\mathrm{SD}}[a,g]$ provides a twistorial description of the SD sector of Yang-Mills theory. That such a description exists is hardly surprising; it is nothing more than a dynamical implementation of the Ward correspondence. What is remarkable is that we can now give a twistorial description of the ASD interactions, thereby completing a perturbative description of full Yang-Mills theory on twistor space. From \eqref{ChS1}, these ASD interactions on space-time are generated by
\be\label{ASDints}
I[G]=\int \d^{4}x\,\tr\!\left(G_{\alpha\beta}\,G^{\alpha\beta}\right)\,.
\ee
To translate this term into twistor data, we simply need to apply the non-abelian integral formula \eqref{nabPT2}:
\begin{multline}\label{IntTA1}
I[a,g]=\int \d^{4}X\,\la\lambda_{1}\,\lambda_{2}\ra^{2}\,\la\lambda_1\,\d\lambda_{1}\ra\,\la\lambda_{2}\,\d\lambda_{2}\ra \\
\times\tr\!\left[\gamma^{-1}(x,\lambda_1)\,g|_{X_1}\,\gamma(x,\lambda_1)\,\gamma^{-1}(x,\lambda_2)\,g|_{X_2}\,\gamma(x,\lambda_2)\right]\,.
\end{multline}
This integral is over two copies (labeled by subscripts 1,2) of the \emph{same} line $X$ in $\PT$, followed by a integration over the four-dimensional moduli space of these lines. This latter integration requires a choice of reality structure on $\PT$ to single out which lines are integrated over; we will assume that the Euclidean reality conditions have been chosen. Note that this action depends implicitly on $a$ through the holomorphic trivialization $\gamma$.

This non-local interaction term can be made to look a bit more twistorial by using the Euclidean reality conditions. With these reality conditions, you can show that the holomorphic volume measure on twistor space is given by:
\be\label{Eucvol}
\D^{3}Z=\la\lambda\,\d\lambda\ra\wedge\lambda_{\alpha}\lambda_{\beta}\,\d x^{\alpha\dot\alpha}\wedge\d x^{\beta}_{\dot\alpha}\,,
\ee
in keeping with the fact that $\PT\cong\R^4\times\CP^1$. This enables us to re-write \eqref{IntTA1} as
\be\label{IntTA2}
I[a,g]=\int\limits_{\PT\times_{\R^4}\PT}\!\!\D^{3}Z_1 \wedge\D^{3}Z_{2}\,\tr\!\left[\gamma^{-1}(x,\lambda_1)\,g(Z_1)\,\gamma(x,\lambda_1)\,\gamma^{-1}(x,\lambda_2)\,g(Z_2)\,\gamma(x,\lambda_2)\right]\,.
\ee
Here, the integral is over the fibre-wise (over $\R^4$) product of two copies of twistor space, each with coordinates $Z^{A}_{1,2}=(x^{\beta\dot\alpha}\lambda_{1,2\,\beta},\,\lambda_{1,2\,\alpha})$.

This leads to a proposal for the full twistor action:
\be\label{TAct}
S[a,g]=S^{\mathrm{SD}}[a,g]+\frac{g^2}{4}\,I[a,g]\,.
\ee
Although it's clear that this must correspond to the space-time action \eqref{ChS1} -- at least in some sense -- by construction, the correspondence between the two is in fact extremely precise~\cite{Mason:2005zm}. The twistor action \eqref{TAct} is literally equal to the space-time action in a particular choice of gauge (one which reduces the remaining gauge freedom to that of space-time gauge transformations), and there is a one-to-one correspondence between extrema of the twistor and space-time actions, with the values of the two functionals agreeing at extrema. In other words, the twistor action is classically equivalent to the space-time action. 

A similar construction can be used to build twistor actions for supersymmetric Yang-Mills theories, all of which admit a similar perturbative expansion around the SD sector~\cite{Boels:2006ir}. Unsurprisingly, the most elegant of these is for the maximal amount of supersymmetry, $\cN=4$; in this case all the degrees of freedom can be packaged into a single twistor field~\cite{Ferber:1977qx}. The twistor action can also be understood from the (equivalent) perspective of `Lorentz harmonic chiral superspace'~\cite{Chicherin:2016fac}, which may be something you have already encountered without knowing that it was related to twistor theory. 

\medskip

Having demonstrated that the googly problem can be overcome perturbatively, one could ask whether the twistor action is actually good for anything. The answer lies in the gauge invariance of the twistor action. A gauge transformation $\gamma(Z)$ on twistor space is a function of three complex variables, or six real variables. Compare this to gauge theory on space-time, where a gauge transformation is a function of only four real variables. So there is a substantially greater functional freedom in the gauge transformations available on twistor space. 

The upshot of this is that there are gauges available on twistor space which are not readily accessible on space-time. Over the last decade, this basic fact has been exploited to derive or prove a wide variety of interesting results in perturbative Yang-Mills theory. A few examples include:
\begin{itemize}
\item Derivation of alternative Feynman rules for Yang-Mills theory, known as `MHV rules'~\cite{Cachazo:2004kj} which substantially simplify the perturbative expansion of physical observables (such as scattering amplitudes)~\cite{Boels:2007qn,Adamo:2011cb}.

\item All-loop integrand expressions for the scattering amplitudes of planar $\cN=4$ super-Yang-Mills theory~\cite{Bullimore:2010pj}.

\item Proof of the scattering amplitudes/Wilson loop duality~\cite{Mason:2010yk,Bullimore:2011ni}.

\item Proof of various correspondences between Wilson loops and limits of correlation functions~\cite{Adamo:2011dq,Adamo:2011cd,Chicherin:2014uca,Koster:2016ebi,Koster:2016fna}.
\end{itemize}
It should be noted that in the case of the latter two examples, these dualities or correspondences were first conjectured using space-time methods or holography (c.f., \cite{Alday:2007hr,Alday:2010zy,Eden:2011yp,Eden:2011ku,Alday:2011ga,Engelund:2011fg}). Although these `traditional' methods generated substantial evidence in favour of the conjectures at both strong and weak coupling, the only known analytic proofs are provided by the twistor action!

\subsubsection*{Exercise: \textit{the non-abelian Penrose transform}}

Working in Euclidean reality conditions, let $\bar{D}=\dbar+a$ be an integrable partial connection on twistor space corresponding to a SD gauge connection on $\R^4$, with a holomorphic trivialization over every $X\cong\CP^1$ given by $\gamma(x,\lambda)$. Show that the integral formulae
\be\label{nabPTe1}
\phi_{\alpha_1\cdots\alpha_{2|h|}}(x)=\int_{X}\la\lambda\,\d\lambda\ra\wedge\lambda_{\alpha_1}\cdots\lambda_{\alpha_{2|h|}}\,\gamma^{-1}(x,\lambda)\,f|_{X}\,\gamma(x,\lambda)\,, \quad h<0\,,
\ee
\be\label{nabPTe2}
\tilde{\phi}_{\dot\alpha_1 \cdots\dot\alpha_{2h}}(x)=\int_{X}\la\lambda\,\d\lambda\ra\wedge\frac{\partial}{\partial\mu^{\dot\alpha_{1}}}\cdots\frac{\partial}{\partial\mu^{\dot\alpha_{2h}}}\,\gamma^{-1}(x,\lambda)\,f|_{X}\,\gamma(x,\lambda)\,, \quad h>0\,,
\ee
define space-time fields which satisfy the covariant z.r.m. equations
\be\label{covzrm}
D^{\alpha_{1}\dot\alpha}\phi_{\alpha_1\cdots\alpha_{2|h|}}=0\,, \qquad D^{\alpha\dot\alpha_{1}} \tilde{\phi}_{\dot\alpha_1 \cdots\dot\alpha_{2h}}=0\,,
\ee
provided that 
\begin{equation*}
 f\in H^{0,1}_{\bar{D}}(\PT,\,\cO(2h-2)\otimes\mathfrak{g})\,.
\end{equation*}

%% file: TLecture5.tex
\section{Beyond Four Dimensions}

Over the last four lectures, we've seen that twistor theory is a useful tool for describing massless free fields and integrable systems (such as the instanton sector) in four-dimensional Minkowski space. We even saw that it was possible to formulate perturbative gauge theory in twistor space. Hopefully, this has convinced you that twistor theory is good for something!

However, it's fair to say that twistor theory -- as we've described it -- still has many shortcomings. The ability to describe massive QFTs remains outside the reach of twistor methods, though this could be overcome using something called the 2-twistor description of massive particles (c.f., \cite{Perjes:1974ra,Penrose:1977in,Hughston:1981zc,Arkani-Hamed:2017jhn}). For massless QFTs, twistor variables have enabled perturbative calculations of loop integrands in planar gauge theories, but actually performing the resulting loop integrations in twistor variables has proved quite difficult (though not impossible, see~\cite{Lipstein:2012vs}). This is due primarily to the non-locality of the relationship between twistor space and space-time as well as the fact that that standard techniques such as dimensional regularization are hard to implement in twistor variables. 

Although we were able to provide a perturbative solution to the googly problem, this will not capture the many physically interesting non-perturbative phenomena which occur in interesting QFTs such as Yang-Mills theory. Even restricting our attention to perturbative QFT, there are many interesting massless theories which still do not have satisfactory descriptions in terms of twistor actions. For instance, conformal gravity -- a conformally invariant, non-unitary theory of gravity which nonetheless has many interesting properties -- has a well-defined twistor action~\cite{Mason:2005zm,Adamo:2013tja}. Yet although general relativity can be classically embedded into conformal gravity~\cite{Maldacena:2011mk}, and the self-dual sector of general relativity has a twistor action~\cite{Mason:2007ct}, it has not yet been possible to extend this to a full perturbative description of Einstein gravity (see~\cite{Herfray:2016qvg} for a survey of various attempts in this direction and their shortcomings). 

\medskip

Many of these issues are the subject of on-going work, and in a few years we may not think of them as major problems for twistor theory. In this lecture, we will talk about another obvious shortcoming of twistor theory, for which there are known solutions: the reliance on 4-dimensions. 

It should be clear by now that the twistor formalism we've been using in these lectures relies intrinsically on space-time being 4-dimensional: otherwise, we can't split vector indices into 2-spinor indices, which is the foundation for everything we've been doing. Though some people might interpret this preference for 4-dimensions as a positive feature of twistor theory, it is difficult to see how to make the formalism useful for interesting topics in higher numbers of dimensions. Fortunately, there are generalizations of the basic concepts of twistor theory beyond 4-dimensions which have proven themselves to be extremely useful in the study of perturbative QFT!


\subsection{From twistors to ambitwistors}

Let's start with the obvious question: can we even define a notion of twistor space for $\M_{\C}$ in dimension $d>4$? The answer is yes, although the definition is a bit technical: $\PT$ is defined to be the space of projective, pure spinors of the complexified conformal group, SO$(d+2,\C)$. A pure spinor is a spinor which obeys some quadratic constraints, the precise form of which are determined by the Clifford algebra in a given dimension. The space of projective pure spinors is simply the space of spinors satisfying these quadratic constraints, considered up to an overall projective scaling.

You might wonder if this $d$-dimensional definition of a twistor is consistent with the $d=4$ formalism we've been using. It's clear that 4d twistors $Z^{A}$ carry an SL$(4,\C)\cong$ SO$(6,\C)$ spinor index which is treated projectively, but we didn't seem to run into any quadratic `purity' constraints. This is because all spinors of SL$(4,\C)$ are automatically pure. As the space-time dimension increases, the purity condition starts to grow teeth, though.

For example, consider $d=6$. In this case $\M_{\C}\cong\C^{6}$ can be charted with complex coordinates $x^{AB}$, where $A,B=1,\ldots,4$ and $x^{AB}=-x^{BA}$ (note these are \emph{not} projective coordinates). The complexified Minkowski metric is given in these coordinates by
\be\label{6dt1}
\d s^{2}=\frac{1}{2}\,\epsilon_{ABCD}\,\d x^{AB}\,\d x^{CD}\,,
\ee
and the corresponding conformal group is SO$(8,\C)$. Just as $\CP^3$ carried a linear action of $\SL(4,\C)\cong\mathrm{SO}(6,\C)$ in 4d, it's clear that $\CP^7$ will carry a natural linear action of SO$(8,\C)$. So a twistor in $d=6$ will be a homogeneous coordinate $\cZ^{I}$ on $\CP^7$, with $I=1,\ldots,8$ considered up to overall projective rescalings.

We still have the purity condition to worry about though; in $d=6$ this amounts to a single quadratic constraint on $\cZ^I$. This can be expressed rather nicely if we split $\cZ^I$ into a twistor and dual twistor coordinate: $\cZ^{I}=(Z^{A}, W_{B})$. In these variables the purity condition is simply $Z^{A}W_{A}=Z\cdot W=0$. Therefore, 6d twistor space takes the form of a projective quadric in $\CP^7$:
\be\label{6dt2}
\PT_{6\d}=\left\{(Z^{A},W_{B})\in\CP^7 | Z\cdot W=0\right\}\,.
\ee
It is straightforward to investigate the geometry of the twistor correspondence in 6d, see~\cite{Mason:2011nw,Saemann:2011nb}. As you might expect, the relationship between $\PT_{6\d}$ and Minkowski space remains non-local, but the dimensionality on either side of the correspondence is enhanced. For instance, a point in $\M_{\C}$ corresponds to a $\CP^3$ inside of twistor space. Similar constructions hold for Minkowski spaces of increasingly higher even dimension~\cite{Baston:1989}, and these also induce natural twistor spaces on odd-dimensional anti-de Sitter space~\cite{Bailey:1998,Adamo:2016rtr}. The general structure is always that of a projective quadric, thanks to the nature of the pure spinor constraints which arise.

Unfortunately, the utility of these higher-dimensional twistor constructions seems to be quite limited in comparison to the 4d case. Although there is a notion of Penrose transform for symmetric spinor fields, these do not correspond to integer-spin z.r.m. fields as they do in $d=4$. Further, non-linear constructions such as the Ward correspondence do not seem to encode non-trivial field configurations as easily as they do in 4d. For example, the Ward correspondence in $d=6$ relates holomorphic vector bundles over $\PT_{6\d}$ to flat gauge fields on space-time~\cite{Baston:1989}. This is due to the intrinsic chirality of the twistor construction: in 4d, there are interesting non-linear gauge field configurations which are chiral (i.e., instantons), but in higher dimensions this is not the case.\footnote{There \emph{are} interesting chiral field configurations in 6d for structures known as \emph{gerbes}. Heuristically, these are like gauge connections, but where the gauge potential 1-form is replaced by a 2-form; a precise definition in the non-abelian case is rather involved. Since the field strength of a gerbe is a 3-form, there are self-dual gerbe in 6d, and these play an important role in the infamous $(2,0)$ superconformal field theory. There is a notion of Ward correspondence for these SD gerbes~\cite{Saemann:2012uq}, but it requires some heavy-duty mathematics (e.g., higher category theory) to set up.} Finally, the quadric constraints appearing in the definition of these higher-dimensional twistor spaces become increasingly byzantine, making it difficult to use the formalism to perform interesting calculations, though twistors have been used to study aspects of QFTs and string theory in higher-dimensions (e.g. \cite{Hughston:1987km,Berkovits:2004bw,Uvarov:2007vs,Bandos:2014lja}).

\medskip

At this point, a pessimist might conclude that twistor theory simply won't be a useful tool beyond 4-dimensions. But we are optimists, so instead of giving up we can try to look for some other construction which mimics the non-locality of the twistor correspondence between Minkowski space and an auxiliary projective space but is non-chiral. Thankfully, such a construction exists, and is known as \emph{ambitwistor theory}~\cite{Witten:1978xx,Isenberg:1978kk,LeBrun:1983}.

Consider complexified Minkowski space $\M_\C$ for any dimension $d$. Let $(X^{a}, P_{b})$ be coordinates on $T^{*}\M_\C$, the cotangent bundle of $\M_\C$. This means that you should think of $X^a$ as a coordinate labeling a point in $\M_\C$, while $P_b$ is a covector specifying a direction at this point. The space of \emph{null directions} in $\M_\C$ is a subspace of this cotangent bundle, given by:
\be\label{nulldir}
T^{*}_{N}=\left\{(X,P)\in T^{*}\M_{\C} | P^2=0\right\}\,.
\ee
We can obtain the space of (complexified) null geodesics in $\M_\C$ by quotienting $T^{*}_{N}$ by shifts up and down each null direction. These shifts are generated by the the vector field $P^{a}\frac{\partial}{\partial X^a}$, so the space of null geodesics is simply
\be\label{nullgeos}
\A=T^{*}_{N}/\left\{P\cdot\frac{\partial}{\partial X}\right\}\,.
\ee
Finally, we can quotient by the scale of each null geodesic to obtain \emph{ambitwistor space},
\be\label{AT1}
\PT=\A/\left\{P\cdot\frac{\partial}{\partial P}\right\}\,,
\ee
which is simply the space of null geodesics in $\M_\C$, up to scale.

Ambitwistor space has many similarities with twistor space: it is a complex projective space (since the quotient by the complex scale of the null geodesics acts as a projective scaling) and is related to space-time non-locally by a double fibration. But unlike twistor space, the ambitwistor correspondence scales uniformly with space-time dimension. Indeed, in $d$ space-time dimensions, ambitwistor space has complex dimensions $2d-3$, and the double fibration is given by:
\begin{equation*}
\xymatrix{
 & \P T^{*}_{N} \ar[ld]_{\pi_{2}} \ar[rd]^{\pi_{1}} & \\
 \PA & & \M_{\C}} 
\end{equation*}
where
\be\label{projnds}
\P T^{*}_{N}=\left\{(X,P)\in T^{*}\M_{\C} | P^2=0\right\}/\left\{P\cdot\frac{\partial}{\partial P}\right\}\,,
\ee
is the space of null directions up to scale. This space always has the topology $\P T^{*}_{N}\cong\M_{\C}\times Q^{d-2}_{\P}$, where $Q^{d-2}_{\P}$ is the space of complexified null directions at a point in $\M_{\C}$. Geometrically, this means that $Q^{d-2}_{\P}$ is a $(d-2)$-dimensional projective quadric. For instance, in $d=4$, it follows that
\begin{equation*}
 Q^{2}_{\P}\cong S^{2}\times S^{2}\cong\CP^{1}\times\CP^1\,,
\end{equation*}
which is the complexification of the space of null directions at a point in Lorentzian-real $\M$ (i.e., the celestial 2-sphere). The fibres of $\pi_{1}:\P T^{*}_{N}\rightarrow \M_{\C}$ are the projective quadrics $Q^{d-2}_{\P}$, while the fibres of $\pi_{2}:\P T^{*}_{N}\rightarrow \PA$ are un-scaled complex null geodesics.

A crucial difference from the twistor construction is that this ambitwistor correspondence easily generalizes when we replace $\M_{\C}$ by \emph{any} complexified space-time, $\cM$. If $g_{ab}$ is the complexified metric on $\cM$, then we can define the space of null directions up to scale by
\be\label{gnulldir}
\P T^{*}_{N}=\left\{(X,P)\in T^{*}\cM\, |\, g^{ab}\,P_{a} P_{b}=0\right\}/\left\{P\cdot\frac{\partial}{\partial P}\right\}\,,
\ee
and ambitwistor space by
\be\label{gAT1}
\PA=\P T^{*}_{N}/D_{0}\,,
\ee
where $D_0$ is the vector field generating the flow along null geodesics in $\cM$:
\be\label{gnullgeos}
D_{0}=g^{ac}\,P_{c}\left(\frac{\partial}{\partial X^{a}}+\Gamma^{d}_{ab}\,P_{d}\,\frac{\partial}{\partial P_{b}}\right)\,.
\ee
The double fibration trivially generalizes to
\begin{equation*}
\xymatrix{
 & \P T^{*}_{N} \ar[ld]_{\pi_{2}} \ar[rd]^{\pi_{1}} & \\
 \PA & & \cM} 
\end{equation*}
so we will just assume that we are working on a generic $d$-dimensional complexified space-time $\cM$ until further notice.

\medskip

The basic correspondence between $\PA$ and space-time is clearly non-local in nature: a point in $\cM$ corresponds to a projective quadric $Q^{d-2}_{\P}\subset\PA$, while a point in $\PA$ corresponds to a complex null geodesic (considered up to scale) in $\cM$. The natural projective scale on $\PA$ is given by assigning projective weight $+1$ to $P$, since we obtain $\PA$ from $\A$ after quotienting by the scale of $P$. This means that there is a natural line bundle over $\mathscr{L}\rightarrow\PA$ given by the functions on $\PA$ which are homogeneous of weight $+1$ in $P$. In our previous notation for line bundles of homogeneous functions, we would say that $\mathscr{L}\cong\cO_{P}(1)$, where the subscript reminds us that this denotes homogeneity in $P$.

Now, the cotangent bundle $T^{*}\cM$ comes with a natural geometric structure, known as a symplectic form: $\omega=\d P_{a}\wedge\d X^{a}$. Here $\omega$ is easily seen to be a non-degenerate and closed 2-form on $T^{*}\cM$. It is also easy to see that $\omega$ arises naturally from a 1-form `symplectic potential' $\theta=P_{a}\d X^{a}$, by
\be\label{atg1}
\theta=P\cdot\frac{\partial}{\partial P}\lrcorner\, \omega\,,
\ee
where $P\cdot\frac{\partial}{\partial P}\lrcorner\,\omega$ denotes the inner product between vectors and differential forms.

If you've been exposed to any symplectic geometry, you will know that every differentiable function $\cF$ on a symplectic manifold determines a vector field on that manifold, known as the \emph{Hamiltonian vector field}, $V_{\cF}$ through the relation:
\be\label{hamvect}
\d\cF=V_{\cF}\lrcorner\, \omega\,,
\ee
where $V\lrcorner\,\omega$ denotes the inner product between vectors and differential forms. Consider the function $-\frac{1}{2} g^{ab}P_{a} P_{b}$ on $T^{*}\cM$; by definition, this vanishes upon restriction to the space of null directions $T^{*}_{N}$. The Hamiltonian vector field of this function on $T^{*}\cM$ is precisely $D_{0}$, the generator of the flow along null geodesics \eqref{gnullgeos}. That is, we have:
\be\label{atg2}
D_{0}\lrcorner\,\omega +\frac{1}{2}\,\d\left(g^{ab}\,P_{a}P_{b}\right)=\Gamma^{c}_{ab}\,P^{a}P_{c}\,\d X^{b}-\frac{1}{2}\left(\Gamma^{a}_{cd}\,g^{db}+\Gamma^{b}_{cd}\,g^{da}\right)P_{a}P_{b}\,\d X^{c}=0\,.
\ee
In terms of the symplectic potential $\theta$, this implies that
\be\label{atg3}
\cL_{D_0}\theta-\frac{1}{2}\,\d\left(g^{ab}\,P_{a}P_{b}\right)=0\,,
\ee
where $\cL_{D_0}$ is the Lie derivative along $D_0$. Upon restriction to $T^{*}_{N}$, this means that $\cL_{D_0}\theta=0$, or that $\theta$ is preserved along the flow of null geodesics. This means that $\theta$ is well-defined on $\PA$.

Thus, the natural geometric structure on ambitwistor space is a holomorphic 1-form $\theta$, inherited from the symplectic structure on $T^{*}\cM$. Since $\theta$ is homogeneous in $P$ of weight $+1$, it is natural to think of it as valued in the line bundle $\mathscr{L}\rightarrow\PA$:
\be\label{contact}
\theta\in\Omega^{1}(\PA,\,\mathscr{L})\,.
\ee
One can show that $\theta$ obeys a non-degeneracy condition:
\be\label{atg4}
\theta\wedge(\d\theta)^{d-2}\neq 0\,.
\ee
Such a 1-form is said to define a (weighted) non-degenerate \emph{contact structure} on $\PA$. A contact structure can be thought of as an odd-dimensional analogue of a symplectic structure (as our derivation of $\theta$ for $\PA$ from the contact structure on $T^{*}\cM$ suggests), and it encodes a substantial amount of interesting geometry. From our perspective, the contact structure $\theta$ on $\PA$ plays the role that the complex structure played on twistor space: it encodes something about the space-time geometry. Indeed, it can be shown that there is an equivalence between the data $(\PA,\theta)$ and the space-time $\cM$ with its torsion-free conformal structure~\cite{LeBrun:1983}.


\subsection{The Penrose transform}

Since the natural geometric structure on $\PA$ is the contact 1-form $\theta$, it makes sense to consider small deformations of the contact structure. In twistor space, cohomological representatives for the Penrose transform can be interpreted as small deformations of the complex structure, and we saw that these led to solutions to free field equations on space-time. Perhaps deformations of the contact structure on $\PA$ will also lead to something interesting on space-time.

We want to consider a deformation $\theta\rightarrow\theta+\delta\theta$, where $\delta\theta$ is sufficiently `small.' In order to get something non-trivial, we have to put some restrictions on this $\delta\theta$; it turns out that the appropriate conditions are: $\delta\theta$ is a $(0,1)$-form on $\PA$, valued in $\mathscr{L}$, which obeys $\dbar\delta\theta=0$, for $\dbar=\d\bar{X}\cdot\frac{\partial}{\partial\bar{X}} +\d\bar{P}\cdot\frac{\partial}{\partial\bar{P}}$ the natural complex structure on $\PA$.\footnote{Such conditions ensure that $\delta\theta$ defines a deformation of the contact structure up to infinitesimal diffeomorphisms.} Furthermore, it can be shown that $\delta\theta$ is a trivial deformation if it can be written as $\delta\theta=\dbar f$ for some function $f$ taking values in $\mathscr{L}$. This means that a non-trivial deformation of the contact structure is a cohomology class:
\be\label{aPT1}
\delta\theta\in H^{0,1}(\PA,\,\mathscr{L})\,.
\ee
Our task is to understand what such a $\delta\theta$ corresponds to on space-time.

First, consider the pullback $\pi_{2}^{*}\delta\theta$ of the deformation to the projective space of null directions, $\P T^{*}_{N}$; this object will be valued in $H^{0,1}(\P T^{*}_{N},\mathscr{L})$. Now, we know that the projective space of null directions is a Cartesian product: $\P T^{*}_{N}\cong\cM\times Q^{d-2}_{\P}$. It turns out that this fact can be used to split the cohomology of $\P T^{*}_{N}$ into cohomology on $\cM$ and $Q^{d-2}_{\P}$, thanks to an important result in homological algebra called the K\"unneth theorem. In the case at hand, this means that
\be\label{aPT2}
H^{0,1}(\P T^{*}_{N},\mathscr{L})\cong H^{0}(\cM)\otimes H^{0,1}(Q^{d-2}_{\P},\,\mathscr{L}) \bigoplus H^{1}(\cM)\otimes H^{0}(Q^{d-2}_{\P},\,\mathscr{L})\,.
\ee
If we assume that $\cM$ has sufficiently boring topology (e.g., that it is topologically equivalent to flat space-time), then it follows that $H^{1}(\cM)=\emptyset$. Furthermore, it can be proved (although we will not show the details here) that the first cohomology of the $(d-2)$-dimensional projective quadrics with values in $\mathscr{L}$ is also trivial: $H^{0,1}(Q_{\P}^{d-2},\mathscr{L})=\emptyset$.

Thus, the K\"unneth decomposition \eqref{aPT2} implies that $H^{0,1}(\P T^{*}_{N},\mathscr{L})=\emptyset$, so we can write
\be\label{aPT3}
\pi^{*}_{2}\delta\theta = \dbar j\,,
\ee
for some $j\in\Omega^{0}(\P T^{*}_{N},\mathscr{L})$. Now, since $\delta\theta$ started life as a cohomology class defined on $\PA$, we must have that $\cL_{D_0}\pi_{2}^{*}\delta\theta =0$. Using Cartan's formula for the Lie derivative of a differential form, this is
\be\label{aPT4}
\cL_{D_0}\pi_{2}^{*}\delta\theta=D_{0}\lrcorner\,\d\left(\pi^{*}_{2}\delta\theta\right) + \d \left(D_{0}\lrcorner\,\pi^{*}_{2}\delta\theta\right)\,.
\ee
But since $\pi^{*}_{2}\delta\theta$ is a $(0,1)$-form cohomology class and $D_0$ is a holomorphic vector field, $D_{0}\lrcorner\,\pi^{*}\delta\theta=0$ and the only contribution comes from the inner product between $D_0$ and the form degrees arising from the exterior derivative in the first term of \eqref{aPT4}. This means that we can write the constraint $\cL_{D_0}\pi_{2}^{*}\delta\theta =0$ as
\be\label{aPT5}
D_{0}\,\pi^{*}_{2}\delta\theta=D_{0}\,(\dbar j) = 0\,,
\ee
using \eqref{aPT3}, where the action of $D_{0}$ is just that of a differential operator. Again using that $D_0$ is a holomorphic vector field, $[D_0,\dbar]=0$, indicating that the constraint \eqref{aPT5} is equivalent to
\be\label{aPT6}
\dbar\,(D_{0}j)=0\,,
\ee
namely, that $D_{0}j$ is holomorphic on $\P T^{*}_{N}$.

From \eqref{gnullgeos}, we see that $D_0$ is homogeneous of weight $+1$ in $P$, which means that \eqref{aPT6} is telling us that
\be\label{aPT7}
D_{0}j\in H^{0}(\P T^{*}_{N},\,\mathscr{L}^{2})\,.
\ee
The usual arguments for homogeneous holomorphic functions therefore indicate that
\be\label{aPT8}
D_{0}j=h(X,P)=h^{ab}(X)\,P_{a} P_{b}\,,
\ee
for some symmetric, trace-free tensor $h_{ab}$ on space-time. Such an $h_{ab}$ is a linear metric perturbation on $\cM$. Using identical arguments, you can show that if we'd started with a trivial deformation (i.e., $\delta\theta=\dbar f$) then the resulting metric perturbation obtained on $\cM$ is pure diffeomorphism: $h_{ab}=\nabla_{(a}\xi_{b)}$ for some $\xi_{b}(X)$.

Thus, we have a statement for the Penrose transform on ambitwistor space:
\be\label{aPTT}
\left\{\mbox{metric perturbations } h_{ab}(X) \mbox{ on } \cM\right\}/\left\{h_{ab}=\nabla_{(a}\xi_{b)}\right\} \cong H^{0,1}(\PA,\,\mathscr{L})\,.
\ee
You can easily generalize this statement to fields of alternative spin by taking cohomology classes on $\PA$ valued in different powers of the line bundle $\mathscr{L}$. Indeed, for integer $n\geq-1$ the Penrose transform reads:
\be\label{aPTT*}
\left\{\mbox{linear fields } \phi_{(a_1 \cdots a_{n+1})_{0}}(X) \mbox{ on } \cM\right\}/\left\{\phi_{(a_1 \cdots a_{n+1})_0}=\nabla_{(a_1}\xi_{a_2\cdots a_{n+1})_0}\right\} \cong H^{0,1}(\PA,\,\mathscr{L}^{n})\,,
\ee
where $\phi_{(a_1 \cdots a_{n+1})_{0}}$ indicates that $\phi_{a_1 \cdots a_{n+1}}$ is totally symmetric and trace-free in its indices.

At first, it might seem that the ambitwistor Penrose transform is actually \emph{more} powerful than the version we learned in twistor space: it makes sense in any dimension and on any complexified space-time. Unfortunately, there is a major shortcoming: the space-time fields generated by the ambitwistor Penrose transform do not obey any equations of motion! Indeed, as we saw in \eqref{aPT8}, the metric perturbation $h_{ab}$ resulting from a deformation of the ambitwistor contact structure is unconstrained (aside from being symmetric and traceless). On twistor space, cohomological data was translated into space-time fields that obeyed free field equations (namely, the z.r.m. equations). We don't seem to get any such equations of motion from the ambitwistor version of the transform.

Considerable effort was put towards trying to find a way to impose field equations through the ambitwistor Penrose transform in the early days of the subject. While it turns out that this can be done, it requires the rather cumbersome formalism of formal neighborhoods~\cite{Witten:1978xx,Isenberg:1978kk,Baston:1987av,LeBrun:1991jh}. In words, this means that equations of motion can be imposed on the resulting space-time fields by demanding that the ambitwistor cohomology representatives on the RHS of \eqref{aPTT*} extend away from the $P^2=0$ quadric to some given order. The major drawback of such a formalism is that it is very difficult to work with; indeed, this led to a dearth of progress in the study of ambitwistor theory until quite recently, when a new strategy for obtaining field equations from the Penrose transform was discovered.

\medskip

Before moving on to these exciting new developments, let's first work through an instructive example of the ambitwistor Penrose transform to ensure that we see exactly what is going on. Take space-time to be $d$-dimensional complexified Minkowski space, $\cM=\M_{\C}$, and consider a plane wave perturbation to the Minkowski metric. This takes the form $h_{ab}=\epsilon_{ab}\,\e^{\im k\cdot X}$, where $\epsilon_{ab}$ is a constant, symmetric and traceless polarization tensor, and $k_{a}$ is a constant $d$-dimensional momentum. This perturbation obeys the linearized Einstein equations if $k^2=0$ and $k^{a}\epsilon_{ab}=0$, but we will see that we can construct the corresponding $\delta\theta$ on $\PA$ without ever needing to impose these conditions.

From $h_{ab}$, we can form
\be\label{apw1}
h(X,P)=\epsilon^{ab}\,\e^{\im\,k\cdot X}\,P_{a}P_{b}\in H^{0}(\P T^{*}_{N},\,\mathscr{L}^2)\,,
\ee
and this must be expressible as $D_{0} j$ for some $j$ taking values in $\mathscr{L}$. Sure enough, it is straightforward to show that:
\be\label{apw2}
j=D^{-1}_{0}\,h=\frac{P_{a} P_{b}}{k\cdot P}\,\epsilon^{ab}\,\e^{\im\,k\cdot X}\,,
\ee
which has the appropriate weight $+1$ in $P$. From \eqref{aPT3}, we can construct the corresponding deformation of the contact structure:
\be\label{apw3}
\pi^{*}_{2}\delta\theta=\dbar j=\bar{\delta}(k\cdot P)\,\epsilon^{ab}\,P_{a}P_{b}\,\e^{\im\,k\cdot X}\,,
\ee
with the holomorphic delta function defined as in \eqref{holdelta}. 

On the support of $k\cdot P=0$, it follows that $D_{0}\pi^{*}_{2}\delta\theta=0$, so \eqref{apw3} descends to $\PA$. Clearly, the resulting $\delta\theta$ is a $(0,1)$-form on with values in $\mathscr{L}$, and it also obeys $\dbar\delta\theta=0$. Note that none of these facts -- or any step in the process of constructing $\delta\theta$ -- requires the linearized Einstein equations.


\subsection{Ambitwistor strings}

The question of how to obtain field equations (even linear ones) from ambitwistor theory in a practical way has a truly remarkable answer: we must combine ambitwistor theory with the 2d conformal field theory (CFT) techniques of \emph{string theory}~\cite{Mason:2013sva}. The motivation for this discovery originated in a series of compact expressions for all tree-level scattering amplitudes in a variety of massless QFTs~\cite{Cachazo:2013hca,Cachazo:2014xea}, but we will simply proceed by looking for a string theory governing maps from a closed Riemann surface $\Sigma$ to ambitwistor space.

Fix space-time to be $d$-dimensional $\M_{\C}$ for simplicity, and let $F:\Sigma\rightarrow\PA$ be a map from the string worldsheet $\Sigma$ to ambitwistor space. What sort of properties should this map have? Well, a recurrent theme throughout these lectures has been holomorphicity, and this applies to ambitwistor space too: we were able to say everything about ambitwistor geometry using only holomorphic coordinates $(X,P)$ on $\PA$. This suggests that a string theory governing $F$ should be holomorphic, or chiral, in nature. 

A natural candidate theory which has this property is one whose kinetic term is the (holomorphic) pullback of the contact structure $\theta$ to the worldsheet:
\be\label{ats1}
S=\frac{1}{2\,\pi}\int_{\Sigma} F^{*}(\theta)-\frac{e}{2}\,P^2 = \frac{1}{2\,\pi}\int_{\Sigma}P_{a}\,\dbar X^{a}-\frac{e}{2} P^2\,.
\ee
Here, $\dbar=\d \bar{z}\,\partial_{\bar{z}}$ is the complex structure on $\Sigma$ in terms of some local affine coordinates $(z,\bar{z})$, while $e$ is a Lagrange multiplier enforcing the quadratic constraint $P^2=0$ necessary for the target space to be $\PA$. This means that the coordinates $(X,P)$ on the target space carry different conformal weight when viewed as fields on $\Sigma$. 

If $X^{a}(z)$ is simply a function on $\Sigma$, then $\dbar X^{a}$ is a $(0,1)$-form, so in order for this worldsheet action to make sense, $P_{a}(z)$ must be a $(1,0)$-form on $\Sigma$. This means that locally, $P_{a}(z)=P_{a\,z}\d z$. In the terminology of 2d CFT, we say that $X^{a}$ has conformal weight $(0,0)$ and $P_{a}$ has conformal weight $(1,0)$ as fields on $\Sigma$. Likewise, the Lagrange multiplier $e$ must have conformal weight $(-1,1)$ in order for the second term in \eqref{ats1} to make sense; locally, this means that $e$ looks like:
\begin{equation*}
 e=e^{z}_{\bar{z}}\,\frac{\d\bar{z}}{\d z}\,.
\end{equation*}
You may have encountered such objects before; they are known as \emph{Beltrami differentials}.

\emph{A priori}, this worldsheet action has $T^{*}_{N}$ as its target space, thanks to constraint $P^2=0$ enforced by the Lagrange multiplier $e$. However, you can check that the action \eqref{ats1} is invariant under the transformations
\be\label{holodiffs}
\delta X^{a}=v\,\partial X^{a}\,, \qquad \delta P_{a}=\partial(v\,P_{a})\,, \qquad \delta e = v\,\partial e-e\,\partial v\,,
\ee
where $v$ is an infinitesimal transformation parameter of conformal weight $(-1,0)$ and $\partial=\d z\,\partial_{z}$. These transformations are infinitesimal holomorphic reparametrizations of the worldsheet $\Sigma$, so the fact that the worldsheet model is invariant under them means that \eqref{ats1} is a classical (holomorphic) 2d CFT. Now, under a holomorphic reparametrization $z\mapsto f(z)$, it follows that the components of $P_{a}$ transform as:
\be\label{hdiffsp}
P_{a\,z}\rightarrow \frac{\partial f}{\partial z}\,P_{a\,f(z)}\,.
\ee
This means that $P_{\mu}$ is only defined up to rescalings by a constant factor, which reduces the target space to $\P T^{*}_{N}$.

But \eqref{holodiffs} are not the only transformations which preserve the worldsheet action. There are also gauge transformations associated with the constraint $P^{2}=0$, under which \eqref{ats1} is invariant:
\be\label{atgtran}
\delta X^{a}=\alpha\,P^{a}\,, \qquad \delta P_{a}=0\,, \qquad \delta e= \dbar\alpha\,,
\ee
for $\alpha$ another infinitesimal gauge parameter of conformal weight $(-1,0)$. Since $P^2=0$, this means that $X^{a}$ is defined only up to translations along any null direction. This is precisely the action of $D_{0}$ in Minkowski space, so the target space of \eqref{ats1} is indeed $\PA$.

\medskip

To quantize this `ambitwistor string theory', we must gauge fix the holomorphic reparametrization invariance and gauge transformations of \eqref{holodiffs} and \eqref{atgtran}, respectively. This can be accomplished with the standard Fadeev-Popov procedure; if we gauge fix to $e=0$ and conformal gauge, then the resulting action is
\be\label{ats2}
S=\frac{1}{2\,\pi}\int_{\Sigma}P_{a}\,\dbar X^{a}+b\,\dbar c+\tilde{b}\,\dbar \tilde{c}\,,
\ee
where $c$, $b$ are the ghost and anti-ghost fields associated with holomorphic reparametrizations, and $\tilde{c}$, $\tilde{b}$ are the ghost and anti-ghost fields associated with the gauge freedom \eqref{atgtran}. All four of these fields have fermionic statistics, and $c,\tilde{c}$ have conformal weight $(-1,0)$ while $b,\tilde{b}$ have conformal weight $(2,0)$. The gauge-fixing also results in a BRST charge given by:
\be\label{BRST}
Q=\oint c\,T+bc\,\partial c + \frac{\tilde{c}}{2}\,P^2\,,
\ee
with 
\be\label{stress}
T=-P_{a}\partial X^{a}-2\,b\,\partial c-\partial b\,c-2\tilde{b}\,\partial\tilde{c}-\partial\tilde{b}\,\tilde{c}\,,
\ee
the holomorphic stress tensor of the worldsheet theory, and normal-ordering assumed for all terms.

Our gauge fixing is anomaly free provided that this BRST charge is nilpotent: $Q^2=0$. This can be checked explicitly by using the free worldsheet OPEs defined by the gauge-fixed action \eqref{ats2}:
\be\label{OPEs}
X^{a}(z)\,P_{b}(w)\sim\frac{\delta^{a}_{b}}{z-w}\,, \qquad c(z)\,b(w)\sim \frac{1}{z-w}\sim\tilde{c}(z)\,\tilde{b}(w)\,.
\ee
You should try this calculation for yourself (it's a chiral version of the famous critical dimension calculation in ordinary string theory); the result is:
\be\label{anom1}
Q^2=\frac{(d-26)}{6}\,c\,\partial^{3}c\,,
\ee
so only the gauge-fixing of the holomorphic reparametrizations is potentially anomalous. The anomaly is fixed by the holomorphic central charge of the fields appearing in the gauge fixed action \eqref{ats2}, and is eliminated with the choice of critical space-time dimension $d=26$.

\medskip

Now, vertex operators in string theories correspond to deformations of the gauge-fixed worldsheet action which are annihilated by the BRST charge. In our case, the interesting part of the action is precisely the contact structure $\theta$ of ambitwistor space, pulled back to the worldsheet. So vertex operators will be given by deformations $\delta\theta$:
\be\label{vero1}
U=\int_{\Sigma} F^{*}(\delta\theta)\,.
\ee
We know, thanks to the Penrose transform, that such $\delta\theta$ correspond to metric perturbations on space-time. Indeed, we can work explicitly with a plane wave deformation \eqref{apw3}, for which the vertex operator takes the form:
\be\label{vero2}
U=\int_{\Sigma}\bar{\delta}(k\cdot P(z))\,\epsilon^{ab}\,P_{a}(z)\,P_{b}(z)\,\e^{\im\,k\cdot X(z)}\,.
\ee
In order for this to be an admissible vertex operator, it must be normal-ordered and obey $Q U=0$.

It is easy to see that these conditions impose further constraints on \eqref{vero2}. Normal-ordering requires that $k_{a}\epsilon^{ab}=0$, while $Q U=0$ if and only if $k^2=0$. This latter constraint comes about from the $P^2$ term in \eqref{BRST}; this is the only part of the BRST charge which has a potentially anomalous contraction with $U$.

But $k_{a}\epsilon^{ab}=0=k^2$ are precisely the linearized Einstein equations for $h_{ab}=\epsilon_{ab}\e^{\im k\cdot X}$! In other words, quantum consistency conditions in the ambitwistor string theory have done what the classical Penrose transform could not: impose linearized field equations on the metric perturbation corresponding to $\delta\theta$. This fact can also be extended to the non-linear level by coupling an ambitwistor string worldsheet model (related to \eqref{ats1} by the addition of some worldsheet fermions) to a non-trivial background metric; quantum consistency of the resulting worldsheet model imposes the non-linear vacuum Einstein equations on this metric~\cite{Adamo:2014wea} (c.f., \cite{Adamo:2015gsa} for a heuristic explanation).

\medskip

The perspective of unifying ambitwistor theory with string methods has led to many exciting advances in recent years. There are far too many examples to mention here in any detail, but one particularly exciting one is related to the calculation of loop corrections to scattering amplitudes in massless QFTs. It turns out that when $\Sigma\cong\CP^1$, correlators of vertex operators in ambitwistor string theories are equal to tree-level scattering amplitudes in a variety of massless QFTs~\cite{Casali:2015vta}. By considering correlation functions on higher genus worldsheets, we can obtain new representations for loop amplitudes~\cite{Adamo:2013tsa,Adamo:2015hoa}! 

Although these higher genus expression are too functionally complicated (involving a localization problem in terms of elliptic functions) to be of practical use from the perspective of a particle physicist, they can be reduced to more manageable expressions by degenerating the underlying Riemann surface into a nodal sphere~\cite{Geyer:2015bja}. This perspective has already led to novel representations of 1- and 2-loop scattering amplitudes in gauge theory and gravity~\cite{Geyer:2015jch,Geyer:2016wjx,Geyer:2017ela}, and looks to be a promising route to obtaining useful new expressions for perturbative amplitudes more generally.

\subsubsection*{Exercise: \textit{the scattering equations}}

\begin{enumerate}
 \item Consider $n-3$ insertions of the vertex operators $U$ given by \eqref{vero2}, and 3 insertions of the `fixed' vertex operators
 \be\label{fixedvo}
 V(z)=c(z)\,\tilde{c}(z)\,\epsilon^{ab}\,P_{a}(z)\,P_{b}(z)\,\e^{\im\,k\cdot X(z)}\,,
 \ee
 in the worldsheet correlation function
 \be\label{wscf}
 \left\la V_{1}(z_1)\,V_{2}(z_2)\,V_{3}(z_3)\,\prod_{i=4}^{n}U_{i}\right\ra\,,
 \ee
 defined by the (Euclidean) path integral with respect to the gauge-fixed action \eqref{ats2}. Show that the path integral over the worldsheet fields $X^{a}(z)$ can be performed explicitly, and that the non-zero-mode portion of this integral enforces the equation 
 \be\label{peqn}
 \dbar P_{a}(z)= 2\pi\im\,\d z\wedge\d\bar{z}\,\sum_{i=1}^{n}k_{i\,a}\,\delta^{2}(z-z_i)\,,
 \ee
 where the $\{z_i\}$ are the $n$ vertex operator insertion points. What is the result of the zero-mode portion of the $X^{a}$ path integral?
 
 \item Solve the equation \eqref{peqn} when $\Sigma\cong\CP^1$. Show that the solution can be written in terms of homogeneous coordinates $\sigma^{\mathrm{a}}=(\sigma^{1},\sigma^{2})$ on the Riemann sphere as
 \be\label{seq1}
 P_{a}(\sigma)=(\sigma\,\d\sigma)\,\sum_{i=1}^{n}\frac{k_{i\,a}\,(i\,p)}{(\sigma\,i)\,(\sigma\,p)}\,,
 \ee
 where $(i\,j):=\sigma^{\mathrm{a}}_{i}\,\sigma^{\mathrm{b}}_{j}\,\epsilon_{\mathrm{ba}}$ is the $\SL(2,\C)$-invariant inner product on these homogeneous coordinates, and $\sigma_p\in\CP^1$ is some auxiliary point. Prove that this solution is independent of the choice of $\sigma_p$. 
 
 \item Compute the quadratic differential $P^{2}(\sigma)$ on $\Sigma\cong\CP^1$, and show that it has only simple poles at the $n$ vertex operator insertion points. Show that the residue of the pole at $\sigma_i$ is given by:
 \be\label{seq2}
 \mathrm{Res}_{\sigma=\sigma_i} P^{2}(\sigma)= (\sigma_{i}\,\d\sigma_{i})\,\sum_{j\neq i}\frac{k_{i}\cdot k_{j}\,(j\,p)}{(i\,j)\,(i\,p)}\,.
 \ee
 
 \item Demonstrate that the remaining ingredients of the worldsheet correlation function \eqref{wscf} enforce
 \be\label{seq3}
 \mathrm{Res}_{\sigma=\sigma_i} P^2(\sigma)=0\,, \qquad i=4,\ldots,n\,.
 \ee
 Why is this equivalent to $\mathrm{Res}_{\sigma=\sigma_i}P^{2}(\sigma)=0$ for \emph{all} $i=1,\ldots,n$? This set of constraints is known as the \emph{scattering equations}.
 
 \item For $P^{2}(\sigma)$ \emph{any} quadratic differential on $\CP^1$ with $n$ simple poles, prove that setting $n-3$ of the residues of these poles equal to zero forces $P^{2}(\sigma)=0$ globally on $\CP^1$ (Hint: use homogeneous coordinates.) We conclude, therefore, that the scattering equations are equivalent to the constraint $P^{2}(\sigma)=0$ -- crucial for the target space of the worldsheet theory to be $\PA$ -- in the presence of vertex operator insertions.
 
\end{enumerate}